\documentclass[useAMS,usenatbib,usegraphicx,a4]{mn2e}
\usepackage{deluxetable} 
\usepackage{graphicx}
\usepackage{verbatim,amsmath,color}
\usepackage{times}
\newcommand{\kms}{km~s$^{-1}$}

\newcommand\solmass{$\rm M_{\sun}$}

\newcommand\peryr{~yr$^{-1}$}
\newcommand{\asec}{\arcsec}
\newcommand{\error}{$\pm$}

\newcommand{\htoo}{H$_2$}
\newcommand{\mhtoo}{M(H$_2$)}

\newcommand{\mhtoomstar}{M(H$_2$)/M$_\star$}
\newcommand{\mhimstar}{M({\sc H\,i})/M$_\star$}
\newcommand{\mhimjam}{M({\sc H\,i})/M$_{\text{JAM}}$}
\newcommand{\mhtoomjam}{M(H$_2$)/M$_{\text{JAM}}$}
\newcommand{\mjam}{M$_{\text{JAM}}$}
\newcommand{\mstar}{M$_\star$}
\newcommand{\logmstar}{\log(M_\star/M_\odot)} 
\newcommand{\logmjam}{\log(M_{\text{JAM}}/M_\odot)} 

\newcommand{\atlas}{{\sc Atlas$^\mathrm{3D}$}}
\newcommand{\e}[1]{$\times 10^{#1}$}
\newcommand{\hbeta}{H$\beta$}
\newcommand{\percc}{~cm$^{-3}$}
\newcommand{\nuvk}{NUV$-$K}
\newcommand{\hi}{{\sc H\,i}}

\begin{document}

\title[Cold Gas in Early-type Galaxies]{The \atlas\ project -- XXVII. Cold Gas and
the Colours and Ages of Early-type Galaxies}
\author[Lisa M. Young et al.]{Lisa M.
Young$^{1,2}$\thanks{E-mail:\texttt{lyoung@physics.nmt.edu}},
Nicholas Scott$^{3}$,
Paolo Serra$^{4,5}$,
Katherine Alatalo$^{6}$,
Estelle Bayet$^{7}$,
\newauthor
Leo Blitz$^{8}$,
Maxime Bois$^{9}$,
Fr\'ed\'eric Bournaud$^{10}$,
Martin Bureau$^{7}$,
Alison F.\ Crocker$^{11}$,
\newauthor
Michele Cappellari$^{7}$,
Roger L. Davies$^{7}$,
Timothy A. Davis$^{12}$,
P. T. de Zeeuw$^{12,13}$,
\newauthor
Pierre-Alain Duc$^{10}$,
Eric Emsellem$^{12,14}$,
Sadegh Khochfar$^{15}$,
Davor Krajnovi\'c$^{16}$,
\newauthor
Harald Kuntschner$^{12}$,
Richard M. McDermid$^{17,18,19}$,
Raffaella Morganti$^{4,20}$,
Thorsten Naab$^{21}$,
\newauthor
Tom Oosterloo$^{4,20}$,
Marc Sarzi$^{22}$,
and Anne-Marie Weijmans$^{23}$\\ 
$^{1}$Physics Department, New Mexico Institute of Mining and Technology, Socorro, NM 87801, USA\\
$^{2}$Academia Sinica Institute of Astronomy \& Astrophysics, PO Box 23-141, Taipei 10617, Taiwan, R.O.C.\\
$^{3}$Sydney Institue for Astronomy (SIfA), School of Physics, The University of
Sydney, NSW 2006, Australia\\
$^{4}$Netherlands Institute for Radio Astronomy (ASTRON), Postbus 2, 7990 AA Dwingeloo, The Netherlands\\
$^{5}$CSIRO Astronomy and Space Science, Australia Telescope National Facility, PO
Box 76, Epping, NSW 1710, Australia\\
$^{6}$Infrared Processing and Analysis Center, California Institute of Technology, Pasadena, California 91125, USA\\
$^{7}$Sub-Dept. of Astrophysics, Dept. of Physics, University of Oxford, Denys Wilkinson Building, Keble Road, Oxford, OX1 3RH, UK\\
$^{8}$Department of Astronomy, Hearst Field Annex, University of California, Berkeley, CA 94720, USA\\
$^{9}$Observatoire de Paris, LERMA and CNRS, 61 Av. de l'Observatoire, F-75014 Paris, France\\
$^{10}$Laboratoire AIM Paris-Saclay, CEA/IRFU/SAp -- CNRS -- Universit\'e Paris Diderot, 91191 Gif-sur-Yvette Cedex, France\\
$^{11}$Ritter Astrophysical Observatory, University of Toledo, Toledo, OH 43606,
USA\\
$^{12}$European Southern Observatory, Karl-Schwarzschild-Str. 2, 85748 Garching, Germany\\
$^{13}$Sterrewacht Leiden, Leiden University, Postbus 9513, 2300 RA Leiden, the Netherlands\\
$^{14}$Universit\'e Lyon 1, Observatoire de Lyon, Centre de Recherche Astrophysique de Lyon and Ecole Normale Sup\'erieure de Lyon, \\
\hspace{3em} 9 avenue Charles Andr\'e, F-69230 Saint-Genis Laval, France\\
$^{15}$Royal Observatory Edinburgh, Blackford Hill, Edinburgh, EH9 3HJ, UK\\
$^{16}$Leibniz-Institut f\"ur Astrophysik Potsdam (AIP), An der Sternwarte 16,
D-14482 Potsdam, Germany\\
$^{17}$Australian Astronomical Observatory, PO Box 296, Epping, NSW 1710, Australia\\
$^{18}$Department of Physics and Astronomy, Macquarie University, NSW 2109, Australia\\
$^{19}$Gemini Observatory, Northern Operations Centre, 670 N. A`ohoku Place, Hilo, HI 
96720, USA\\
$^{20}$Kapteyn Astronomical Institute, University of Groningen, Postbus 800, 9700 AV Groningen, The Netherlands\\
$^{21}$Max-Planck-Institut f\"ur Astrophysik, Karl-Schwarzschild-Str. 1, 85741 Garching, Germany\\
$^{22}$Centre for Astrophysics Research, University of Hertfordshire, Hatfield, Herts AL1 9AB, UK\\
$^{23}$School of Physics and Astronomy, University of St Andrews, North Haugh, St
Andrews KY16 9SS, UK\\}

\maketitle

\begin{abstract}
We present a study of the cold gas contents of the \atlas\ early-type
galaxies, in the context of their optical colours, near-UV colours, and 
\hbeta\ absorption line strengths.
Early-type (elliptical and lenticular) galaxies are not as gas-poor as previously
thought, and at least 40\% of local early-type galaxies are now known 
to contain molecular and/or atomic gas.  
This cold gas offers the opportunity to study recent galaxy evolution
through the processes of cold gas acquisition,
consumption (star formation), and removal.
Molecular and atomic gas detection rates range from 
10\% to 34\% in red sequence early-type galaxies, depending on how the red sequence
is defined, 
and from 50\% to 70\% in blue early-type galaxies.
Notably, massive red sequence early-type galaxies  (stellar masses
$ > 5\times 10^{10}$ \solmass, derived from dynamical models) are
found to have \hi\ masses up to \mhimstar~$\sim 0.06$ and \htoo\ masses up to 
\mhtoomstar~$\sim 0.01$.
Some $20\%$
of all massive early-type galaxies may have retained atomic and/or 
molecular gas through their transition to the red sequence.  However,
kinematic and metallicity signatures of external gas accretion (either from
satellite galaxies or the intergalactic medium) are also common,
particularly at stellar masses $\leq 5\times 10^{10}$ \solmass, where such signatures are found in
$\sim 50\%$ of \htoo-rich early-type galaxies.
Our data are thus consistent with a scenario in which fast rotator
early-type galaxies are quenched former spiral galaxies which have undergone some
bulge growth processes, and in addition, some of them also experience cold gas
accretion which can initiate a period of modest star formation activity.
We discuss implications for the interpretation of colour-magnitude diagrams.
\end{abstract}

 
\section{INTRODUCTION}

While early-type (elliptical and lenticular) galaxies generally have smaller
relative
amounts of star formation activity than spirals, they are not all completely
devoid of such activity.
Modest rates of star formation in early-type galaxies can be traced with UV colours,
mid-IR and far-IR continuum emission, mid-IR PAH emission, optical emission lines,
and cm-wave radio continuum emission
\citep[e.g.][]{shapiro,sarzi10}.  
Surveys of UV and optical colours have suggested that this kind of low-level star
formation activity is present in around a quarter of all nearby early-type
galaxies \citep{yi05,kaviraj07,suh10}.  
Recently \citet{smith12} have emphasized that star formation
can be found even in early-type galaxies which are on the red sequence, and 
\citet{fumagalli} have measured star formation rates in
optically quiescent galaxies out to redshifts of 1--2.

This evidence for star formation activity 
should be understood in the context of paradigms for the development of
the red sequence.  Star formation requires cold gas, and the treatment
of cold gas is a crucial element of galaxy evolution models.
For example, one class of models assumes that the development
of the red sequence is driven by AGN activity, as AGN-driven outflows remove the
star-forming gas from blue gas-rich galaxies \citep[eg.][]{kaviraj11}.
The outflow is assumed to remove the gas more quickly than it could be consumed by
star formation activity.
Cold gas may also be removed or consumed by a variety of other 
processes associated with bulge growth \citep{cheung12}.
Another class of models points out that it may not be necessary to {\it remove} 
the cold gas, but simply to sterilize it \citep{kawata,martig,martig12}.
In order to test these paradigms for the development of the red
sequence, and in particular to test the relative importance of 
gas recycling, accretion and removal as galaxies move to and from the red sequence, 
we explore the
relationships between the colours of early-type galaxies, their \hbeta\ absorption
line strengths, and their cold gas
content.

There have been several other surveys for
molecular gas in various samples of early-type galaxies \citep[e.g.][]{WSY10}, but
the \atlas\ project is the only one which also has 
the kinematic information that is necessary to interpret galaxies' recent
evolutionary histories.  Stellar kinematics, shells and tidal
features give insight into the merger and
assembly histories of the galaxies, and ionized gas, \hi, and molecular kinematics
reveal signs of gas accretion and/or gravitational disturbances. 
Thus, the \atlas\ project is the first opportunity to 
bring together the evidence from star formation activity (recorded
in the colours and \hbeta\ absorption of the stellar populations) and the 
recent interaction/accretion history recorded in the gas.
In this paper we do not deal with measurements of an instantaneous star formation
rate or star formation efficiencies derived therefrom
\citep[e.g.][]{martig12}.
Instead, we
focus particularly on the connections
between gas accretion and the star formation history over the last few 
Gyr, as those are encoded in the stellar populations.

Section \ref{sec:sample} of this paper describes the \atlas\ sample of local early-type
galaxies.  Section \ref{sec:data} gives explanatory information on the provenance
of the colour, stellar population, and cold gas data.
Section \ref{sec:redseq} shows the atomic and molecular gas contents of early-type
galaxies, both on and off the red sequence. 
Depending on which colours are used to define the red sequence, 10\%\error 2\%
to 34\%\error 6\% of
red sequence early-type galaxies have $> 10^7$~\solmass\ of cold gas.  
A discussion of internal extinction is found in Section
\ref{sec:internal}, which shows that although the \htoo-rich galaxies are 
dusty, the dust usually has only modest effects on integrated colours.
Section \ref{sec:accretion} compiles multiple types of evidence for recent gas
accretion and finds no systematic difference in the colours or stellar populations
of galaxies with and without such evidence.  A ``frosting" or rejuvenation model that invokes a small
quantity of recent star formation on a red sequence galaxy can explain the colours and
\hbeta\ absorption line strengths of the \htoo-rich \atlas\ early-type galaxies.
Section \ref{sec:highmass} notes some differences between the high (stellar) mass
\htoo\ detections and their low (stellar) mass counterparts.
Section \ref{sec:summary} summarizes the conclusions of the work.

\section{SAMPLE}\label{sec:sample}

The \atlas\ early-type galaxy sample is a complete volume-limited sample of galaxies brighter than
$M_K = -21.5,$ covering distances out to 42 Mpc, with some
restrictions on Declination and Galactic latitude \citep[see][hereafter Paper I]{paper1}.
The early-type sample is actually drawn from a parent sample which has no
colour or morphological selection, and optical images of the entire parent sample
have been inspected by eye for large-scale spiral structure.  
The 260 galaxies lacking spiral structure form the basis of the \atlas\ project;
integral-field optical spectroscopy over a field of at least 33\asec $\times$
41\asec\ was obtained with the SAURON
instrument \citep{bacon01} on the William Herschel Telescope for these 260 galaxies.
Paper~I also tabulates their assumed distances and foreground optical extinctions.

The \atlas\ early-type galaxies
clearly trace out the optical red sequence, and they also include a smaller population 
of somewhat bluer galaxies (Paper I).
Detailed study of the stellar kinematic maps is used to reveal internal
substructures such as counter-rotating stellar cores or kinematically decoupled
stellar discs \citep[][Paper II]{paper2}.
The specific angular momenta of the galaxies are analyzed in \citet{paper3}
(Paper III) in the context of formation paradigms for slow and fast rotators
\citep{paper6,paper8,naab}.
Dynamical masses are provided by \citet{paper15,paper20}.
Other work on the sample probes environmental drivers of galaxy evolution
\citep{paper7,cappellari-apjl}, hot gas \citep{sarzi},
stellar populations and star formation histories \citep{mcd13}, 
and bulge/disc decompositions \citep{decomp}.

\section{DATA}\label{sec:data}

\subsection{Cold gas, line strengths and stellar population ages}

All of the \atlas\ early-type galaxies except NGC\,4486A
were observed with the IRAM 30m
telescope in the $^{12}$CO J=1-0 and 2-1 lines
\citep[][Paper IV]{paper4}.\footnote{NGC\,4486A was missed because
the velocity information in NED and HyperLEDA was inaccurate at the time of the
obervations.} 
The CO detection limits correspond to \htoo\ masses
of approximately $10^7$ \solmass\ for the nearest galaxies and $10^8$ \solmass\ 
for the more distant galaxies.  
To place this CO survey in context it is crucial to note that the luminosity
selection criterion for the sample is based purely on the total $K_S$-band stellar
luminosity, not the FIR flux or even the $B$ luminosity 
(which have been used in
the past, and which are strongly influenced by
the presence of star formation). 
The CO detection rate is 22\%\error 3\%; detected \htoo\ masses range from $10^{7.1}$ 
to $10^{9.3}$ \solmass\ and the molecular/stellar mass ratios \mhtoomstar\ 
range from 3.4\e{-4} to 0.076.  
In addition, the CO detection
rate and \mhtoomstar\ distributions are broadly consistent between Virgo Cluster
and field galaxies, so that the cluster environment has not strongly affected the
molecular gas content.

The brightest CO detections in the SAURON sample \citep{dz02} were
mapped in CO J=1-0
emission with the BIMA and Plateau de Bure millimeter interferometers,
giving the molecular gas distribution and kinematics at typically 5\asec\
resolution \citep{y02, y05, YBC, crocker2768, crocker4550, crocker-all}.
Similarly, the brightest CO detections in the remainder of the
\atlas\ sample were
mapped with CARMA for this project \citep{DRpaper}.
These CO maps have been obtained for a total of 40 \atlas\ galaxies, which is by far the largest and
most robustly defined currently available sample of molecular gas maps of
early-type galaxies.  The molecular gas is found in kpc-scale structures
\citep{davis13}, often in discs or rings, and these data provide kinematic
information for
our discussion of the galaxies' histories and their location/movement in the
colour-magnitude diagram.

Observations of \hi\ are available for all of the \atlas\ early-type galaxies at declinations $>
10^\circ,$ except the four closest in projection to M87 \citep{paolo}.
The \hi\ detection rate is 32\%\error 4\%,
and \hi\ mass upper limits range from 2.9\e{6} \solmass\ to 3.8\e{7} \solmass.
Detected \hi\ masses fall in roughly the same range as the molecular masses, i.e.
\mhimstar $= 10^{-4}$ to $10^{-1}$.
These interferometric observations are made with the
Westerbork Synthesis Radio Telescope (WSRT), covering $\sim$ 30\arcmin\ fields of
view at typical resolutions of 35\asec $\times$ 45\asec.
The data thus allow us to consider both the total \hi\ content of the galaxy and to
isolate the portion of atomic gas which is located
(in projection) in the central regions of the galaxy, in an area comparable to the
SAURON field of view and about three times larger than the beam of the 30m
telescope.
The difference between the total and central \hi\ measurements can be significant
because, as \citet{O10} and \citet{paolo} have noted, the \hi\ distributions
occasionally extend to more than 10 effective radii; furthermore, \hi\ is sometimes
only detected in the outskirts of a galaxy.  
As we will show, the central \hi\ detections are much better
correlated with molecular gas than are the total \hi\ detections.
The central \hi\ masses are presented in Table \ref{tab:centralhi}.

\begin{table*}
\begin{minipage}{170mm}
\caption{Central \hi\ masses \label{tab:centralhi}}
\begin{tabular}{lclclclclcl}
\hline
{Name} &  & log M(\hi)/\solmass & & Name & & log M(\hi)/\solmass & & 
Name & & log M(\hi)/\solmass \\
\hline
      IC0598 &  $<$ &  7.06       & &   NGC3674 &  $<$ &  7.02       & &   NGC4649 &  $<$ &  7.19       \\
      IC3631 &  $<$ &  7.34       & &   NGC3694 &  $<$ &  7.11       & &   NGC4660 &  $<$ &  6.50       \\
     NGC0661 &  $<$ &  6.99       & &   NGC3757 &  $<$ &  6.72       & &   NGC4694 &   =  &  7.13 (0.02)\\
     NGC0680 &   =  &  7.65 (0.05)& &   NGC3796 &  $<$ &  6.72       & &   NGC4710 &   =  &  6.71 (0.08)\\
     NGC0770 &  $<$ &  7.17       & &   NGC3838 &  $<$ &  7.23       & &   NGC4733 &  $<$ &  7.12       \\
     NGC0821 &  $<$ &  6.53       & &   NGC3941 &  $<$ &  6.17       & &   NGC4754 &  $<$ &  7.18       \\
     NGC1023 &   =  &  6.84 (0.02)& &   NGC3945 &  $<$ &  6.73       & &   NGC4762 &  $<$ &  7.41       \\
     NGC2481 &  $<$ &  7.03       & &   NGC3998 &   =  &  7.42 (0.02)& &   NGC5103 &   =  &  7.30 (0.04)\\
     NGC2549 &  $<$ &  6.12       & &   NGC4026 &  $<$ &  7.14       & &   NGC5173 &   =  &  8.45 (0.01)\\
     NGC2577 &  $<$ &  6.96       & &   NGC4036 &  $<$ &  6.80       & &   NGC5198 &  $<$ &  6.98       \\
     NGC2592 &  $<$ &  6.80       & &   NGC4078 &  $<$ &  7.26       & &   NGC5273 &  $<$ &  6.42       \\
     NGC2594 &  $<$ &  7.22       & &   NGC4111 &   =  &  6.94 (0.04)& &   NGC5308 &  $<$ &  7.24       \\
     NGC2679 &  $<$ &  6.97       & &   NGC4119 &  $<$ &  7.10       & &   NGC5322 &  $<$ &  6.96       \\
     NGC2685 &   =  &  7.36 (0.02)& &   NGC4143 &  $<$ &  6.42       & &   NGC5342 &  $<$ &  7.12       \\
     NGC2764 &   =  &  8.91 (0.01)& &   NGC4150 &   =  &  6.04 (0.06)& &   NGC5353 &  $<$ &  7.07       \\
     NGC2768 &  $<$ &  6.61       & &   NGC4168 &  $<$ &  7.08       & &   NGC5355 &  $<$ &  7.11       \\
     NGC2778 &  $<$ &  6.68       & &   NGC4203 &   =  &  7.03 (0.03)& &   NGC5358 &  $<$ &  7.13       \\
     NGC2824 &   =  &  7.45 (0.08)& &   NGC4251 &  $<$ &  6.58       & &   NGC5379 &  $<$ &  6.97       \\
     NGC2852 &  $<$ &  6.90       & &   NGC4262 &  $<$ &  7.02       & &   NGC5422 &   =  &  7.43 (0.05)\\
     NGC2859 &  $<$ &  6.85       & &   NGC4267 &  $<$ &  7.17       & &   NGC5473 &  $<$ &  7.02       \\
     NGC2880 &  $<$ &  6.65       & &   NGC4278 &   =  &  6.06 (0.09)& &   NGC5475 &  $<$ &  6.89       \\
     NGC2950 &  $<$ &  6.31       & &   NGC4283 &  $<$ &  5.97    &  &     NGC5481 &  $<$ &  6.83       \\
     NGC3032 &   =  &  7.80 (0.01)& &   NGC4340 &  $<$ &  6.65    &  &     NGC5485 &  $<$ &  6.79       \\
     NGC3073 &   =  &  8.01 (0.02)& &   NGC4346 &  $<$ &  6.27    &  &     NGC5500 &  $<$ &  6.97       \\
     NGC3098 &  $<$ &  6.73       & &   NGC4350 &  $<$ &  6.50    &  &     NGC5557 &  $<$ &  7.16       \\
     NGC3182 &   =  &  6.93 (0.16)& &   NGC4371 &  $<$ &  7.10    &  &     NGC5582 &  $<$ &  6.88       \\
     NGC3193 &  $<$ &  7.07       & &   NGC4374 &  $<$ &  6.88    &  &     NGC5611 &  $<$ &  6.76       \\
     NGC3226 &  $<$ &  6.72       & &   NGC4377 &  $<$ &  7.16    &  &     NGC5631 &   =  &  7.54 (0.03)\\
     NGC3230 &  $<$ &  7.33       & &   NGC4379 &  $<$ &  7.04    &  &     NGC5687 &  $<$ &  6.94       \\
     NGC3245 &  $<$ &  6.61       & &   NGC4382 &  $<$ &  6.59    &  &     NGC5866 &   =  &  6.67 (0.06)\\
     NGC3248 &  $<$ &  6.84       & &   NGC4387 &  $<$ &  6.65    &  &     NGC6149 &  $<$ &  7.18       \\
     NGC3301 &  $<$ &  6.75       & &   NGC4406 &  $<$ &  6.40    &  &     NGC6278 &  $<$ &  7.28       \\
     NGC3377 &  $<$ &  6.14       & &   NGC4425 &  $<$ &  6.33    &  &     NGC6547 &  $<$ &  7.25       \\
     NGC3379 &  $<$ &  6.11       & &   NGC4429 &  $<$ &  7.12    &  &     NGC6548 &  $<$ &  6.74       \\
     NGC3384 &  $<$ &  6.19       & &   NGC4435 &  $<$ &  7.23    &  &     NGC6703 &  $<$ &  6.80       \\
     NGC3400 &  $<$ &  6.81       & &   NGC4452 &  $<$ &  7.27    &  &     NGC6798 &   =  &  8.10 (0.02)\\
     NGC3412 &  $<$ &  6.17       & &   NGC4458 &  $<$ &  6.53    &  &     NGC7280 &   =  &  7.25 (0.05)\\
     NGC3414 &  $<$ &  7.70       & &   NGC4459 &  $<$ &  6.53    &  &     NGC7332 &  $<$ &  6.70       \\
     NGC3457 &   =  &  6.95 (0.07)& &   NGC4461 &  $<$ &  7.33    &  &     NGC7454 &  $<$ &  6.78       \\
     NGC3458 &  $<$ &  6.97       & &   NGC4473 &  $<$ &  6.47    &  &     NGC7457 &  $<$ &  6.22       \\
     NGC3489 &   =  &  6.53 (0.03)& &   NGC4474 &  $<$ &  7.09    &  &     NGC7465 &   =  &  8.64 (0.01)\\
     NGC3499 &   =  &  6.77 (0.14)& &   NGC4477 &  $<$ &  6.56    &  &   PGC028887 &  $<$ &  7.29       \\
     NGC3522 &  $<$ &  7.48       & &   NGC4489 &  $<$ &  6.35    &  &   PGC029321 &  $<$ &  7.30       \\
     NGC3530 &  $<$ &  6.98       & &   NGC4494 &  $<$ &  6.46    &  &   PGC035754 &  $<$ &  7.20       \\
     NGC3595 &  $<$ &  7.04       & &   NGC4503 &  $<$ &  7.15    &  &   PGC044433 &  $<$ &  7.28       \\
     NGC3599 &  $<$ &  6.64       & &   NGC4521 &  $<$ &  7.18    &  &   PGC050395 &  $<$ &  7.13       \\
     NGC3605 &  $<$ &  6.44       & &   NGC4528 &  $<$ &  7.19    &  &   PGC051753 &  $<$ &  7.13       \\
     NGC3607 &  $<$ &  6.53       & &   NGC4550 &  $<$ &  6.50    &  &   PGC061468 &  $<$ &  7.15       \\
     NGC3608 &  $<$ &  6.53       & &   NGC4551 &  $<$ &  7.39    &  &   PGC071531 &  $<$ &  6.98       \\
     NGC3610 &  $<$ &  6.63       & &   NGC4552 &  $<$ &  6.48    &  &    UGC03960 &   =  &  7.06 (0.11)\\
     NGC3613 &  $<$ &  6.90       & &   NGC4564 &  $<$ &  6.53    &  &    UGC04551 &  $<$ &  6.87       \\
     NGC3619 &   =  &  8.25 (0.01)& &   NGC4596 &  $<$ &  7.13    &  &    UGC05408 &   =  &  8.33 (0.02)\\
     NGC3626 &   =  &  7.80 (0.02)& &   NGC4608 &  $<$ &  7.22    &  &    UGC06176 &   =  &  8.40 (0.02)\\
     NGC3648 &  $<$ &  6.99       & &   NGC4621 &  $<$ &  6.48    &  &    UGC08876 &  $<$ &  7.05       \\
     NGC3658 &  $<$ &  7.04       & &   NGC4638 &  $<$ &  7.13    &  &    UGC09519 &   =  &  7.75 (0.02)\\
     NGC3665 &  $<$ &  7.05       & &           &      &             &
&      & \\

\hline
\end{tabular}

\textit{Notes:} Central \hi\ fluxes are those observed within one WSRT resolution
element centred on the optical nucleus \citep[for additional description see][]{O10}.
We attempt to account for beam smearing effects in
that the cases with poorly resolved central \hi\ holes are treated as upper
limits. 
The upper limits are calculated as a 3$\sigma$ uncertainty on a sum
over one beam and 50 \kms; that strategy is motivated by \citet{paolo}.
Uncertainties on \hi\ masses are quoted as 1$\sigma$ on a sum over one beam and 50
\kms, with a 3\% absolute calibration uncertainty added in quadrature.
\end{minipage}
\end{table*}

\citet{mcd13} present \hbeta\ absorption line strength measurements for the
\atlas\ galaxies, along with the best-fit single or simple stellar population
(SSP) ages and metallicities.  As discussed by \citet{trager00}
and \citet{kuntschner10}, for example, the SSP
parameters can be thought of as luminosity-weighted mean stellar properties,
although the weight bias of the SSP ages towards young populations is even 
stronger than their luminosities would suggest \citep{serra+trager}.
The measurements are made in circular apertures of
radius $R_e$ (the galaxy's effective or half-light radius), $R_e/2$, and $R_e/8$.  Here we make use of the $R_e/2$ apertures, on
the grounds that they are most similar to the size of the IRAM 30m beam, and the
$R_e/8$ apertures, which are the most sensitive to small amounts of recent star
formation activity near the galaxy centers.

\subsection{Optical and UV photometry}

The Sloan Digital Sky Survey (SDSS) $u$ and $r$ magnitudes and foreground
extinctions for \atlas\ 
galaxies are obtained from Data Release 8 \citep[DR8,][]{sdss8}.
The matched-aperture `modelMag' magnitudes are used to compute colours and 
the Petrosian $r$ magnitude is used for an absolute magnitude $M_r$.  
For five bright galaxies we were unable to find object matches of any type, primary or
secondary, in DR8, even though primary matches are identified in DR7; for these
five the DR7 magnitudes are used.  They are NGC numbers 3608, 4374, 4459,
4486, and 4649. 
Some caution is required in these cases.  \citet{sdss8} note that the DR8
photometry was computed with an updated sky measurement algorithm, and the update
tends to make extended galaxies brighter.  Indeed, a comparison of DR7 and
DR8 photometry for the \atlas\ galaxies suggests that 
objects of $r \leq 12$ are about 0.5 mag brighter in 
DR8 than in DR7.  Thus the DR7 photometry of the `missing' galaxies 
is manually adjusted 0.5 mag brighter to
make the values commensurate with DR8 photometry.  The offset between DR8 and DR7 is roughly
uniform in all filters and therefore does not distort colours as much as luminosities.
\citet{blanton11} and \citet{scott12} have further shown that the 
DR8 catalog
photometry still underestimates the fluxes of extended galaxies by 0.5 mag or more,
but \citet{blanton11} have again demonstrated that the effect on colours is minimal.

\begin{figure}
\includegraphics[scale=0.6,trim=1cm 0.5cm 1.5cm 0.5cm]{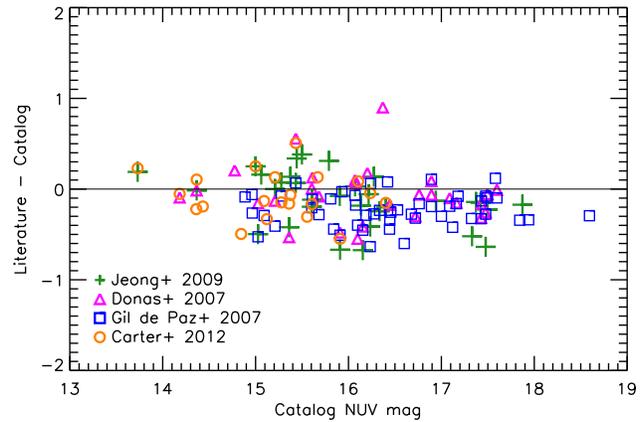}
\caption{Comparison of previously published NUV magnitudes and GALEX catalog GR7
data for the \atlas\ sample.  The pink triangle outlier is NGC\,4578, for which the
new catalog data use a deeper exposure than \citet{donas07} used.
}
\label{fig:galex_comp}
\end{figure}

Near-UV magnitudes for the \atlas\ galaxies were obtained from the 
GALEX catalog server, release GR7.   They are 
corrected for foreground extinction assuming the Milky Way $E(B-V)$ values from
\citet{schlegel}, just as for the SDSS data above, and scaled to NUV as $A_{NUV}
= 8.0*E(B-V)$ \citep{gildepaz07}.  
One quarter
of the \atlas\ galaxies have additional published NUV photometry derived from extrapolation
to curves of growth or from two-dimensional image fits, and Figure
\ref{fig:galex_comp}
presents a comparison of the catalog photometry to the results of \citet{donas07},
\citet{gildepaz07}, \citet{jeong09}, and \citet{carter11}.   There may be a
systematic offset on the order of 0.2 mag for faint galaxies (NUV $\ge 16$), in
the sense that the published magnitudes are brighter than the catalog values.

We use the asymptotic NUV magnitudes  and the 2MASS K$_S$-band magnitudes
\citep{2mass} to
compute \nuvk\ colours.  While this procedure is not as accurate as
aperture matching, it is adequate for our purposes here as we use the photometry
only to indicate the distributions of galaxies within a colour-magnitude diagram.
Likewise any systematic offsets on the order of 0.2 mag are still small compared
to the scatter in the colours of red sequence galaxies.

\subsection{Stellar masses}
Two different estimates of stellar mass are used in this paper.  One, denoted
\mstar, is derived from 
$K_S$ luminosities using $M_{K\odot} =
3.28$ \citep{BT} and a typical
stellar mass-to-light ratio of $M_\star/L_K = 0.82$ in solar units \citep{bell03}.
The other, denoted \mjam\ and used more frequently, is a dynamical mass derived from
self-consistent modeling of the stellar kinematics and the isophotal structure of each
galaxy \citep{paper15}. 
It can be understood as \mjam~$\approx 2\times M_{1/2}$, where $M_{1/2}$ 
is the total mass within a sphere
enclosing half of the galaxy light.  Given that the proportion of dark matter is 
small inside that
sphere, \mjam\ approximates a stellar mass that also accounts for
systematic variations in the initial mass function \citep{paper20}.
The derivation of \mjam\ assumes that the stellar mass-to-light ratio is spatially
constant within a galaxy, but unlike the case of \mstar, it does not assume the ratio is the same for all
galaxies.  Trends in the stellar mass-to-light ratios are discussed in
\citet{paper20}.  For our purposes here, we note that the two measures of stellar mass are 
nearly linearly correlated with the median value of
$\log (M_\text{JAM}/M_\star)$ being 0.17 dex and the dispersion being 0.15 dex.

\section{Cold gas in red sequence galaxies}\label{sec:redseq}

\subsection{Molecular gas}

\begin{figure}
\includegraphics[scale=0.55,trim=1.5cm 0.5cm 0cm 0.5cm]{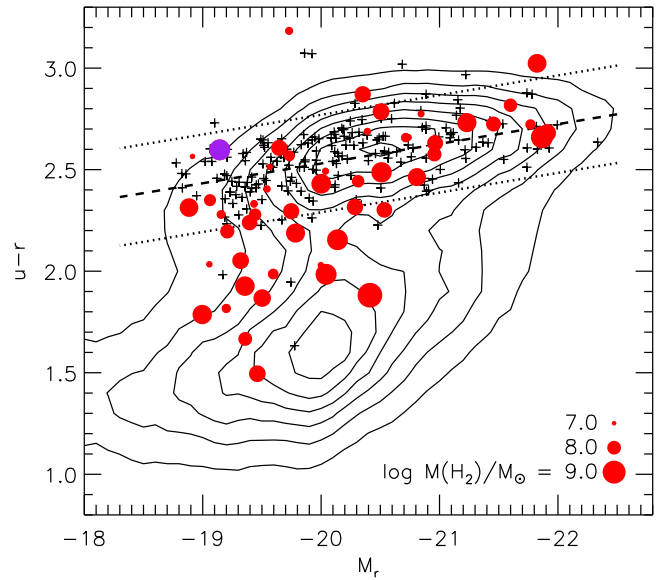}
\caption{Optical colour-magnitude diagram for the \atlas\ sample.  CO detections are 
marked with red
circles and nondetections with black crosses; the sizes of the red circles
indicate the value of \mhtoo, scaled logarithmically as indicated in the
legend.  Contours underneath show the red sequence and the blue cloud as
demarcated by a
sample of 60000 galaxies with redshifts in the range $0.01 \leq z \leq 0.08$ from SDSS Data Release 8.
No $\mathrm{V/V_{max}}$ correction is applied, so the contours
are intended to mark the general locations of the red sequence and the blue cloud in
this diagram but not to indicate relative numbers of galaxies in different regions.
The dashed line is the red sequence ridgeline and the dotted ones are parallel
lines 2$\sigma$ redder and
bluer, as described in section \ref{sec:implications}.
The purple symbol is UGC\,09519, which is discussed in the text along with
NGC\,1266 as examples of galaxies with red colours but strong \hbeta\ absorption.
NGC\,1266 does not appear in this plot because it is outside the SDSS coverage
area.
\label{fig:CMD}
}
\end{figure}

\begin{figure}
\includegraphics[scale=0.55,trim=1.5cm 0.5cm 0cm 0.5cm]{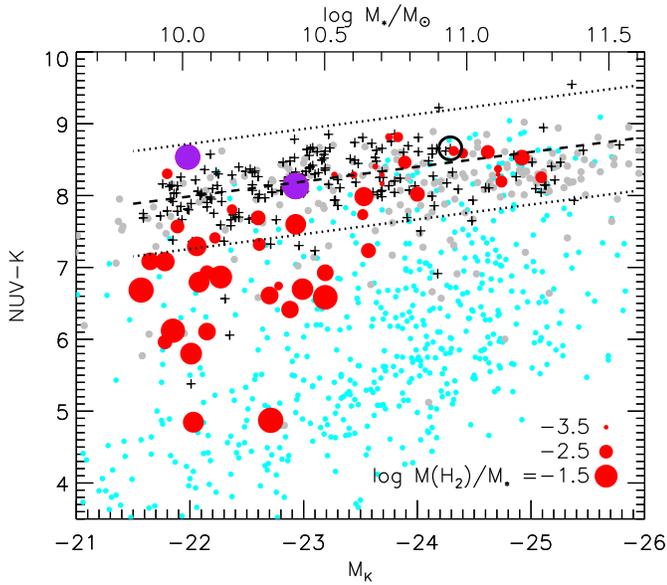}
\caption{\nuvk\ colour-magnitude diagram with \htoo\ masses.
Symbols and dotted lines are as in Figure \ref{fig:CMD}. For context, cyan and
grey dots are
from the GALEX UV Atlas of Nearby Galaxies \citep{gildepaz07}, 
where cyan dots are late type galaxies (morphological type $T\ge0$) and grey dots
are early types ($T<0$).
Note that the NUV magnitudes are measured in the
AB system whereas the $K_S$ magnitudes are in the Vega system.
Since the \atlas\ GALEX data are taken from the GALEX pipeline catalog,
and Figure \ref{fig:galex_comp} shows a median offset of 0.19 mag between the 
pipeline and 
the NUV magnitudes of \citet{gildepaz07}, the data from that paper have been
shifted to account for the relative offset.
Purple symbols mark NGC\,1266 and UGC\,09519, which are discussed in the text;
NGC\,1266 is the higher luminosity one of the pair. 
NGC\,4552, a well-known UV upturn galaxy, is circled in black to demonstrate that
the effects of the UV upturn are comparable to the dispersion in red sequence
colours.
\label{fig:galex_cmd}
}
\end{figure}

\begin{figure}
\includegraphics[scale=0.55,trim=1.5cm 0.5cm 0cm 0.5cm]{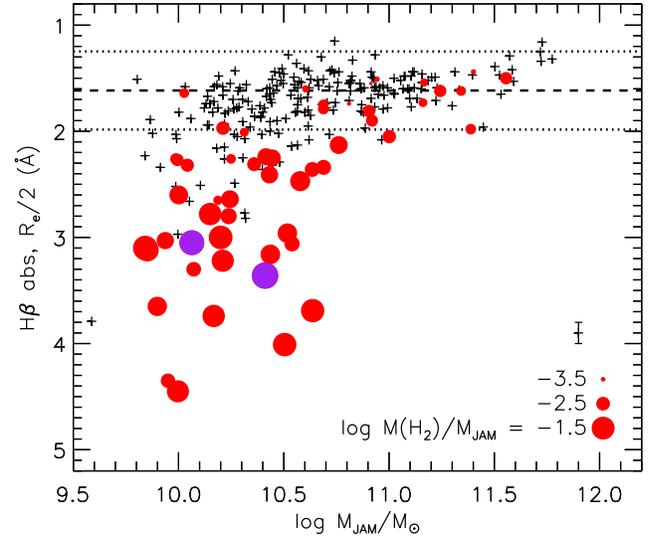}
\caption{\hbeta\ absorption line strength vs.\ 
stellar mass \mjam, with \htoo\ masses.
Symbols and lines are as for Figures \ref{fig:CMD} and \ref{fig:galex_cmd}; purple symbols
again are NGC1266 and UGC09519.  The mean uncertainty in the \hbeta\ line strength
is 0.1 \AA\ \citep{mcd13}, and it is indicated in the lower right corner with an
error bar showing $\pm 1\sigma$.
\label{fig:hbeta_cmd}
}
\end{figure}

\begin{figure}
\includegraphics[scale=0.55,trim=1.5cm 0.5cm 0cm 0.5cm]{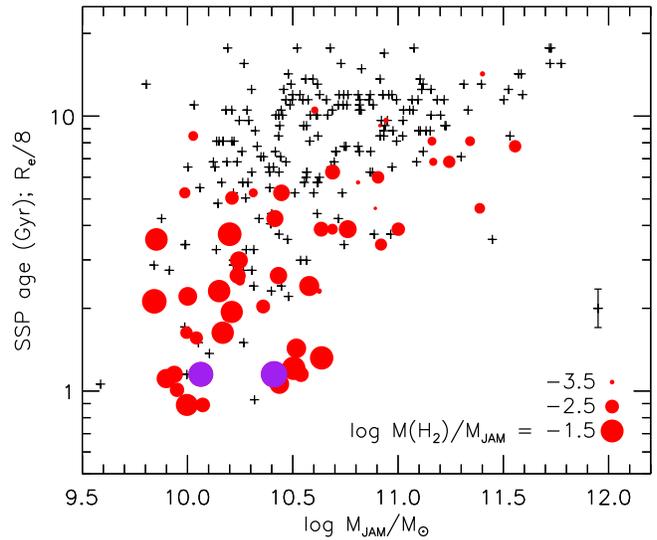}
\caption{``Colour-magnitude" diagram using the inferred stellar SSP ages in place
of colours and dynamical masses \mjam\ in place of absolute magnitude. 
NGC\,1266 and UGC\,09519 are again shown in purple.
The uncertainties in the SSP ages range from $\Delta$(log age) = 0.04
dex to 0.12 dex, with a mean of 0.07 dex \citep{mcd13}, and that uncertainty is
shown in the lower right corner with an
errorbar of $\pm 0.07$ dex.
}
\label{fig:agejam_cmd}
\end{figure}

Figures \ref{fig:CMD} through \ref{fig:agejam_cmd} show 
colour-magnitude diagrams and their analogs for the
\atlas\ early-type galaxies, with symbol sizes scaled to the molecular gas masses.
Figure \ref{fig:CMD} presents $u-r$ colours and
Figure \ref{fig:galex_cmd} shows \nuvk\ colours with the modification that 
molecular masses are normalized to the stellar mass as \mhtoomstar.
A comparison of the two panels provides visual illustration of important 
statistical results from Figure 7 of Paper IV, namely that 
the CO detection rate and \mhtoo\ distributions are surprisingly
constant over the luminosity range of the sample.  
Thus in Figure \ref{fig:galex_cmd} there is a trend for \mhtoomstar\ to be larger for low luminosity galaxies, but
statistically speaking it is because $M_\star$ is smaller and not because \mhtoo\ is
larger.
\citet{paper20} have also shown that, at fixed stellar mass, galaxies
with low stellar velocity dispersion and large effective radius have larger molecular masses;
in other words, molecular gas in early-type galaxies is preferentially associated 
with a discy stellar component.  The masses of the stellar discs in question are
much larger than the currently observable molecular masses, though, as suggested
by the \mhtoomstar\ values in Figure \ref{fig:galex_cmd}.

The red sequence is clearly evident in these figures, as are 
a number of the \atlas\ members in
the green valley and even into the blue cloud. 
These `blue tail' early-type galaxies are found in the lower mass portion of
the sample, with $M_r > -20.5$, 
$M_K > -23.4$ or $\logmjam < 10.7$; according to the luminosity functions
derived by \citet{bell03} they have $L \leq
0.6L^*$. Galaxies with higher masses are still on the red sequence.
The `blue tail' early-type galaxies are therefore analogous to those detected
in clusters at moderate redshift, for example, by \citet{jaffe11}.
The CO detection rate is 0.54 \error\ 0.06 to 0.70 \error\ 0.08 among them, depending
on which colours are used to define the blue tail.  (More extensive discussion of
cold gas detection rates will be  
found in section \ref{sec:implications}.)  
They have the highest values of \mhtoomjam\ in the sample, namely 0.01 to 0.07,
and those values are comparable to the typical spiral galaxies in the COLD GASS
sample \citep{kauffmann13}.
These facts support the 
suggestion that the blue tail galaxies are blue because of ongoing star formation.
They are also known to have younger stellar populations in their centers
\citep{scott12,kuntschner10}.  The relatively few CO-nondetected blue tail
galaxies tend to be at large distances, $D \geq 35$ Mpc, so we suspect their
molecular gas contents are just below our sensitivity limits.
In contrast to the blue tail galaxies, many of the CO detections, particularly
those with $L \ge 0.6 L^*$ or $\logmjam > 10.7$,
also belong to galaxies which
are located in the heart of the optical and UV--NIR red sequence.

Figure \ref{fig:hbeta_cmd} presents an analog of a
colour-magnitude diagram, constructed using the \hbeta\ absorption line strength
index (interior to $R_e/2$)
in the role of colour.  
The dynamical mass \mjam\ is used in place of a stellar
luminosity.
As both the \hbeta\ line strength and \nuvk\ colour are sensitive to the presence of
young stellar populations, and
as the \atlas\ sample displays a tight relationship between
\hbeta\ line strength and \nuvk\ colour, 
Figures \ref{fig:galex_cmd} and \ref{fig:hbeta_cmd}
are qualitatively similar.  The only notable differences are two outliers in
the (\nuvk) -- \hbeta\ relation, NGC\,1266 and UGC\,09519, which are both red sequence
galaxies in \nuvk\ colours but have strong \hbeta\ absorption.  
They might be experiencing a recent shutdown in star formation activity, as an
abrupt
shutdown would show up in the NUV flux before the \hbeta\ line strength \citep[see
also][]{katey1266sf}.
These
galaxies are also known to have internal reddening from dust, and colour images are 
presented in \citet{DRpaper}.

Even more so than the \nuvk\ diagram, the \hbeta-mass diagram
dramatically emphasizes
a dichotomy between a tight red sequence at high stellar masses and 
a large dispersion in \hbeta\ absorption line strengths at lower stellar masses. 
The demarcation line, at around $\logmjam = 10.7,$ also marks
changes in the structural properties of the \atlas\ galaxies.  Above that mass,
the galaxies tend to be spheroid-dominated systems.  Below that mass, early-type
galaxies include both spheroid-dominated and disc-dominated systems with a wide
range of velocity dispersions and effective radii.  The disc-dominated systems
have more molecular gas and younger stellar populations than the spheroid-dominated
systems \citep{paper20}.

Figure \ref{fig:agejam_cmd} involves the highest degree of analysis and
interpretation, as its axes are constructed
from the dynamical
mass \mjam\ and the SSP age \citep{mcd13}.  
Here we have used the SSP age
measured in a central aperture of radius $R_e/8$ (typically only a couple of
arcseconds) and we observe that the smaller apertures are more
sensitive to recent star formation activity than large apertures \citep[see also][]{mcd13}.  For example, the
CO-detected galaxies with \mjam~$> 10^{11}$ \solmass\ do not exhibit enhanced
\hbeta\ absorption in an $R_e/2$ aperture (Figure \ref{fig:hbeta_cmd}), but they do
exhibit younger ages in an $R_e/8$ aperture.
While the trend remains for a smaller 
age dispersion at higher masses, it is not such a sharp disparity as in \nuvk\
colours and \hbeta\ line strength.

Figures \ref{fig:CMD} through \ref{fig:agejam_cmd} give a different perspective on
the \htoo\ content of early-type galaxies than the CO searches of the COLD GASS project \citep{saintonge}. 
That paper finds no CO emission in any red sequence galaxy, but its CO
observations were not as deep and its sensitivity
limits would miss most of the detections in the \atlas\ sample.
The stellar mass range covered by both samples is very similar.
However, only 8 of the 56 CO detections in the \atlas\ sample would simultaneously
meet the stellar mass and CO sensitivity criteria of \citet{saintonge}, and only
two of those (NGC\,1266 and UGC\,09519) are on the red sequence in \nuvk\
and $u-r$ colours.
A stacking analysis of the COLD GASS data is presented in \citet{saintonge3}, but
stacking removes all information about which individual galaxies have gas.
The improvements in sensitivity of the \atlas\ project pick up significant
numbers of molecular gas detections in red sequence early-type galaxies.  
Extreme examples of these red
sequence galaxies with small but nonzero molecular masses can also be found in the 
detections of shocked \htoo\ emission in radio galaxies \citep{ogle}.

\subsection{Atomic gas}\label{atomicgas}

\begin{figure}
\includegraphics[scale=0.53,trim=1cm 0cm 0cm 0cm]{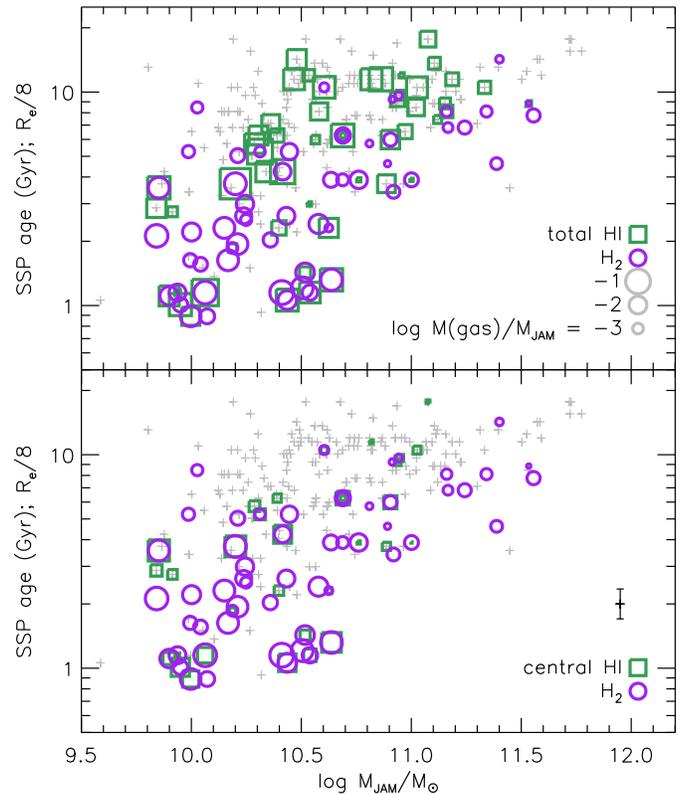}
\caption{Atomic gas \citep{paolo} and molecular gas in the age-mass diagram.
As in the previous figure, the errorbar in the lower right corner
shows a typical uncertainty in the SSP age.
Gray crosses are CO nondetections; cold gas contents are indicated by the sizes of
the green squares (\hi) and purple circles (\htoo). 
The top panel shows the total \hi\ content associated with the galaxy, whereas the
bottom panel shows only the \hi\ content within the central WSRT synthesized beam 
($\sim$ 30\asec\ -- 40\asec\ diameter, roughly comparable to the 22\asec\
aperture used for CO measurements).
The \htoo/\hi\ mass ratios can also be inferred from the symbol sizes; a galaxy
with \htoo/\hi~= 1 has a concentric circle and square of the same diameter.
In comparing atomic and molecular masses, it is worthwhile to remember that
some of the \htoo\ detections lack \hi\ data due to the Declination limit on the
WSRT observations.  However, all of the \hi\ detections do have CO measurements.
\label{fig:hi_cmd}
}
\end{figure}

The \atlas\ sample displays a broad range of \htoo/\hi\ mass ratios,
with measured values from $10^2$ to $10^{-2}$ 
\citep[see also][]{WSY10}.
Figure \ref{fig:hi_cmd} presents the age-mass diagram with \htoo, total \hi\, and
central \hi\ masses.
When considering the total \hi\ contents, the distribution of 
\hi-rich galaxies
is markedly different from that of the \htoo-rich galaxies.  For example,
only 8 of 56 \htoo-rich galaxies (0.14\error 0.05) have SSP ages $\ge$ 8 Gyr; however, 19 of
53 \hi-rich galaxies (0.36\error 0.07) have such large ages.  For space considerations we have 
not shown \hi\ contents on
the optical, NUV or \hbeta-magnitude diagrams, but the same effect is present
there and is even stronger at dynamical masses $\log(M_{\text{JAM}}/M_\odot) <
10.5.$ Interestingly, \htoo\ is primarily
detected in fast rotators \citep{paper4} whereas \hi\ is detected in both fast
and slow rotators \citep{paolo-kin}, so the slow rotators are more commonly
\hi-rich than \htoo-rich. 

When considering only
the central \hi\ masses, however (Figure \ref{fig:hi_cmd}, lower panel), we find that the galaxies with
central \hi\ detections have age and mass distributions indistinguishable from 
those of the \htoo\ detections.
The central \hi\ masses are comparable to or smaller than the \htoo\ masses,
as already noted by \citet{O10} and \citet{paolo}, and this is undoubtedly due to
the fact that the atomic column densities are smaller than the molecular column
densities in the galaxy centers \citep{br06,leroy08,lucero12}.
The one 
known exception with a large (central \hi)/\htoo\ mass ratio $\sim 3.1 \pm 0.6$ is
NGC\,3073, and in the absence of a CO map for the galaxy we can only
speculate that its CO emission may be distributed over a smaller area than its \hi.

These data strengthen the suggestions of \citet{morganti06},
\citet{serra08}, and \citet{O10} that the extended atomic gas 
reservoirs in early-type galaxies are not
necessarily associated with recent star formation activity as traced in the 
galaxies' integrated colours, \hbeta\ absorption line strength indices, or SSP
ages. As \citet{paolo} have noted, the \hi\ column densities in early-type
galaxies tend to be smaller than those in spirals. 
Atomic gas may be associated with star formation activity in an extended UV-bright
disc, as in \citet{thilker10}, but more careful surveys of GALEX data and deep optical imaging
\citep[e.g.][]{megacam1} should be conducted to assess
this activity in the \atlas\ sample.

In summary, molecular gas and {\it central} atomic gas in early-type galaxies
are almost always associated with recent star formation 
activity which can be seen in the SSP ages, particularly at
small radii.
For early-type galaxies with $\logmjam < 10.7$, the recent star formation
activity manifests itself also in the \hbeta\ line strength and \nuvk, but often
it is not noticeable in integrated $u-r$ colours.
For the highest mass early-type galaxies the star formation activity is not
detectable in either optical or UV--NIR integrated colours.

\subsection{Implications of cold gas in red sequence galaxies}\label{sec:implications}

To quantify the cold gas content in red sequence early-type galaxies, we first make an
empirical definition of the red sequence in \atlas.  We fit lines to the
colours and \hbeta\ absorption line strengths of all galaxies not detected in 
CO emission; these lines define the
ridgelines of the red sequence.  In the case of \hbeta\ absorption the fitted line is
constrained to have no slope.
The dispersions characterizing the width of the red sequence are estimated by 
fitting Gaussians to the histograms of residuals (again using only CO
nondetections).  Such a Gaussian fit is more robust to outliers than simply computing
the standard deviation directly.  Galaxies within $2\sigma$
of the ridgeline, or redder, are defined to be red sequence galaxies.  
Figures \ref{fig:CMD} through \ref{fig:hbeta_cmd} show the red sequence ridgelines and
the blue edges defined in this manner.
Table \ref{tab:params} gives the red
sequence parameters and Table \ref{tab:redseq} gives the detection rates for 
cold gas in red sequence galaxies.
In the \atlas\ sample the galaxies with $\logmjam > 10.7$ or $L > 0.6L^*$ are red sequence
galaxies, with very few exceptions, and therefore detection statistics for the high-mass
galaxies are also given.

\begin{table}
\caption{Red Sequence Definitions\label{tab:params}}
\begin{tabular}{lcc}
\tableline
\tableline
{``Colour"} & {Ridgeline} & {$\sigma$} \\
\tableline
$u-r$ & $u-r = 0.59-0.097 (M_r)$ & 0.12 mag\\
\nuvk & \nuvk $= 3.83-0.187 (M_K)$ & 0.37 mag\\
\hbeta~$(R_e/2)$ & \hbeta\ $=1.62$ \AA & 0.18 \AA \\
\tableline
\end{tabular}
\end{table}

\begin{table*}
\begin{minipage}{150mm}
\caption{Cold Gas Detections in Early-Type Galaxies\label{tab:redseq}}
\begin{tabular}{llccccc}
\hline
{Galaxy Subtype} & Definition &
{CO data} & {\hi\ data} & \multicolumn{3}{c}{Detection rates} \\
{} & {} & {} & {} & {\htoo} & {Central \hi} & {Total \hi} \\
{(1)} & {(2)} & {(3)} & {(4)} & {(5)} & {(6)} & {(7)} \\
\hline
all                 & & 259 & 166 & 22 (03) & 19 (03) & 32 (04) \\
$u-r$ red sequence  & $u-r > 0.35-0.097(M_r)$ & 213 & 144 & 16 (02) & 15 (03) & 30 (04)  \\
\nuvk\ red sequence & $NUV-K > 3.10-0.187(M_K)$ & 200 & 127 & 14 (03) & 13 (03) & 27 (04)  \\
\hbeta\ red sequence & \hbeta\ $ < 1.98$ \AA & 189 & 123 & 10 (02) & 09
(03) & 24 (04)  \\
high mass ($\approx$ red sequence)&  $\logmjam > 10.7$  & 102 & 61 & 16 (04) & 15 (04) &
34 (06) \\
low-mass, \hbeta\ red sequence & $\logmjam \le 10.7$ and \hbeta\ $< 1.98$ \AA & 93 & 66 & 05 (02) & 06 (03) & 18 (05) \\
low-mass, \hbeta\ `blue tail' & $\logmjam \le 10.7$ and \hbeta\ $\ge 1.98$ \AA  & 64 & 39 & 55 (06) & 49 (08) & 51 (08) \\
\hline
\end{tabular}
\textit{Notes:} Column 3 gives the total number of \atlas\ galaxies in each category
that have CO data; this is the entire sample, minus NGC\,4486A. 
Column 4 gives the number with \hi\ data \citep{paolo}.  The detection rates in
columns 5--7 are percentages and include
formal uncertainties in parentheses.
\end{minipage}
\end{table*}

We note, first, that the detection rates of cold gas in early-type red sequence galaxies are
still substantial.  For example, in \htoo\ they range from 10\% 
(using \hbeta\ to define the
red sequence) to 16\% ($u-r$), and in total \hi\ content
they are 24\% (using \hbeta) to 34\% (high-mass galaxies). 
This result is qualitatively consistent with the detection rate of dust through 
far-IR emission from red sequence early-type galaxies \citep{smith12}.
When low-mass early-type galaxies are detected in cold gas emission, they tend to be blue; 
however, when high-mass early-type galaxies are similarly detected, they are still on the red
sequence.  Thus, for red sequence early-type galaxies,
the cold gas detection rates
are higher among the more massive galaxies than among the
low-mass galaxies.
The effect can be seen by comparing rows 5 and 6 in Table \ref{tab:redseq}.
Detection rates of (total) atomic gas are also always higher in red sequence
galaxies than detection rates of molecular gas.
In the \atlas\ sample the red sequence early-type galaxies
have total \hi\ masses up to 5\e{9} \solmass, central \hi\ masses up to 2.5\e{8}
\solmass\ and \htoo\ masses up to 2\e{9} \solmass.  Those latter two maximum values
depend at a factor of two level on the indicator (colour or \hbeta\ index) used to
define the red sequence.  Normalized to stellar masses, the red sequence galaxies
have total ratios \mhimjam\ $\leq 0.16$, central \mhimjam\ $\leq
0.01$ and \mhtoomjam\ $\leq 0.07$.

At high stellar masses, the red sequence is just as well-defined 
for galaxies rich in cold gas as for galaxies poor in cold gas.  Specifically, the
Kolmogorov-Smirnov and Mann-Whitney U tests show no statistically significant
differences in the integrated colours or \hbeta\ absorption strengths of the 
high-mass CO
detections and nondetections, or the \hi\ detections and nondetections.
The aperture must be restricted to very small sizes (e.g.\ $R_e/8,$
as in Figure \ref{fig:agejam_cmd}) before enhanced \hbeta\ values become measurable in \htoo-rich massive
galaxies.

Colour-magnitude relations are often used to study the quenching of star formation
and the development of the red sequence \citep[e.g.][]{snyder12,jaffe11}.
Our data show that the approach to the red sequence may not necessarily involve
the loss of all cold gas, even if one makes both colour {\it and} morphological
cuts to select early-type galaxies.
Thus it should not be assumed that gravitational
interactions between such galaxies will always be dissipationless \citep[`dry'; see
also][]{chou12}.  
Furthermore, at $z=0$ we observe that
the presence of \htoo\ or \hi\ does not increase colour scatter in $L > 0.6L^*$
early-type galaxies.
This result could have implications for the so-called scatter-age test
\citep[e.g.][]{jaffe11}, which uses colours to infer star formation activity in high
luminosity galaxies.  For the most massive \atlas\ galaxies, neither colours nor
\hbeta\ absorption line strengths identify
\htoo- or \hi-rich galaxies, though for low luminosities the colours and \hbeta\ can be
partially successful at selecting gas-rich galaxies.  
Finally, if all
galaxies' cold gas contents were generally higher in the past than they are now,
our data at $z=0$ may provide only lower limits on the cold gas contents of red
sequence galaxies at higher redshifts.

\section{Colour corrections for internal extinction}\label{sec:internal}

The previous section analyzed colour-magnitude relations using integrated colours
without correction for internal dust reddening.  Those uncorrected colours and
their dispersions are appropriate for comparison to intermediate and high 
redshift studies of the red sequence 
\citep[e.g.][]{snyder12,jaffe11}, where colours are also typically not corrected for
internal dust.  We now estimate how large such corrections would be.

Many different methods have been used to estimate the internal extinction
in galaxies, including scaling from the far-IR thermal dust emission, 
the Balmer decrement, and/or the
4000 \AA\ break, as well as from detailed fits of the spectra
\citep[e.g.][]{wyder07,schiminovich07,cortese08,goncalves}.
But since early-type galaxies have smooth light distributions, and
dust discs and filaments are readily visible in optical images of the
\atlas\ galaxies \citep{paper2,scott12,DRpaper}, we adopt the more straightforward method of
measuring the internal extinction directly from optical images.

In \citet{scott12}, $g-i$ colour maps are used to identify dusty regions and to
de-redden those regions in $r$ images.  We allow for intrinsic colour gradients
within galaxies by measuring the local $g-i$ colour excess from a linear trend in
$g-i$ with radius \citep[see Figure 2 of][]{scott12}.  The resulting total internal extinctions $A_r$ are measured as
the difference in the integrated magnitude before and after the de-reddening
correction, and are presented in Table \ref{tab:ar}.  The corresponding 
extinctions for $u$, NUV, and K$_S$ are calculated from a $R_V = 3.1$ reddening
law \citep{sf11} and, as above, $A_{NUV} = 8.0 E(B-V)$ \citep{gildepaz07}.
Measured total extinctions $A_r$ range from 0.01 mag to 0.5 mag, with
a median of 0.05 mag.  For context and
comparison, a dusty patch imposing 1 mag of extinction and covering 10\% to 20\% of an 
otherwise uniform source produces a total extinction of 0.067 to 0.139 mag.

Consistency checks on the extinctions in Table \ref{tab:ar} can be made from the 
FIR luminosities of our sample galaxies. 
Specifically, following the prescription of \citet{johnson07},
we compute the expected FUV extinction from the infrared excess
$ \mathrm{IRX} = \log(L_{dust}/L_{UV}).$  The justification of this method is that the FIR
emission is reprocessed from UV emission intercepted by dust grains, so the FIR/UV
ratio is a {\it de facto} measurement of the UV extinction.  The expected FUV
extinction can then be scaled back to $A_r$ by our assumed reddening law.
With one exception (NGC\,2685, which has a dramatically bright FUV disc or ring),
these predicted values of $A_r$ are 3 to 20 times larger than the measured values
of $A_r$.  Indeed, this kind of a discrepancy between IR-derived and
optically-derived dust masses has been known for many years \citep{goudfrooij95},
and it is not clear whether the IRX method can be applied in early-type galaxies
in the same way as it is in spirals \citep{johnson07}.
We therefore adopt the optically-derived extinctions and simply bear in mind that
they may be underestimated, as they will tend to miss smoothly distributed dust
components.
FIR-based dust masses for many of the \atlas\ sample have also been presented by 
\citet{smith12}, \citet{martini}, and \citet{auld}, among others; investigations of
the dust properties are, however, beyond the scope of this paper. 

Internal extinctions have not been measured for every galaxy in the \atlas\
sample.  They have only been measured in cases where the de-reddened images were
necessary for robust two-dimensional image modelling in \citet{scott12}.  Hence
some notable dusty galaxies are not represented in Table \ref{tab:ar}, either
because the dust is spatially compact (e.g.\ UGC\,09519 and NGC\,1266), or a dust
disc is nearly
face-on (e.g.\ NGC\,3032), or the
dust is so optically thick that the de-reddening correction fails (e.g.\
NGC\,4710 and NGC\,5866).  Nevertheless, the set with internal extinction data is an
illustrative set of the gas-rich \atlas\ galaxies.  Of the 24, 19 are detected in
CO emission and an additional three are detected in \hi\ emission (the remaining two 
have no \hi\ data).  \citet{DRpaper}, \citet{crocker-all}, and \citet{young09} show
some overlays of optical colour maps and CO
integrated intensity distributions in the \htoo-rich \atlas\ galaxies, to illustrate the close association of
molecular gas and dust.

\begin{table}
\caption{Internal extinctions\label{tab:ar}}
\begin{tabular}{lrclr}
\hline
{Name} & {$A_r$ (mag)} & {~~~} &
{Name} & {$A_r$ (mag)} \\
\hline
IC1024  &  0.092      & &  NGC4233  &  0.041      \\
NGC2685 &    0.52     & &  NGC4281  &  0.021      \\
NGC2764 &   0.049     & &  NGC4324  &  0.073      \\
NGC3156 &  0.0038     & &  NGC4429  &   0.22      \\
NGC3489 &   0.012     & &  NGC4459  &  0.014      \\
NGC3499 &   0.092     & &  NGC4476  &  0.013      \\
NGC3607 &   0.016     & &  NGC4526  &   0.16      \\
NGC3619 &   0.051     & &  NGC4753  &  0.052      \\
NGC3626 &   0.024     & &  NGC5379  &  0.048      \\
NGC3665 &   0.031     & &  NGC5631  & 0.0075      \\
NGC4036 &    0.13     & &  PGC056772&     0.22      \\
NGC4119 &    0.12     & &  UGC06176 &   0.020    \\
\hline
\end{tabular}
\end{table}

No internal extinction correction is attempted for the \hbeta\ absorption index. 
In principle, as it is a narrow-baseline equivalent width measurement and the
extinction will be approximately constant over a small wavelength range, \hbeta\
is robust to a screen of dust.
In practice, however, the \hbeta\ line strengths could be underestimated (and the
inferred SSP ages could be too old) if
young stars are completely obscured behind dust clouds. 
Similarly, the colour correction measures colour excess with respect to the
stellar populations surrounding the dust clouds or discs.  If the stellar
populations behind the dust clouds are younger, the broadband
extinction corrections could be underestimated.  
This kind of an underestimate is probably a larger effect for the
broadband colour than for the \hbeta\ index as the stars dominating \hbeta\
absorption are older and will have moved farther from their natal clouds.

\begin{figure}
\includegraphics[scale=0.55,trim=1.5cm 0.5cm 0cm 0.5cm]{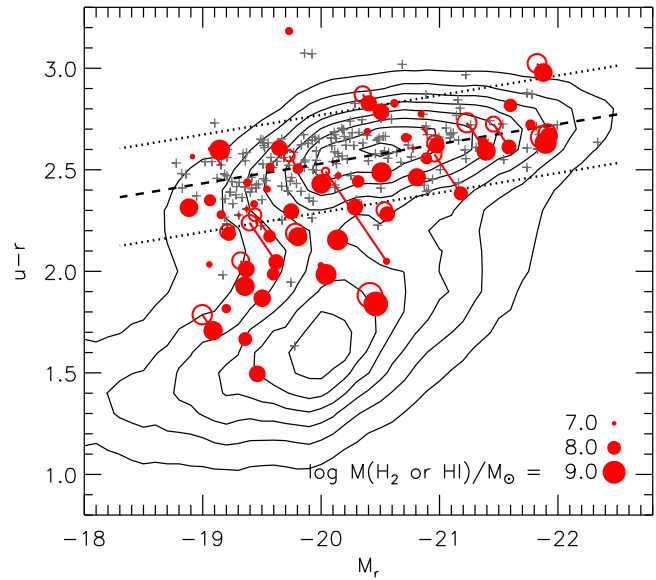}
\caption{Similar to Figure \ref{fig:CMD}, except that galaxies with central \hi\
detections have been added, and corrections for internal 
extinction have been made for the galaxies in Table \ref{tab:ar}.  Corrected
photometry values (filled circles) are linked with a line to the values before correction (open
circles).
\label{fig:CMD_corr}
}
\end{figure}

\begin{figure}
\includegraphics[scale=0.55,trim=1.5cm 0.5cm 0cm 0.5cm]{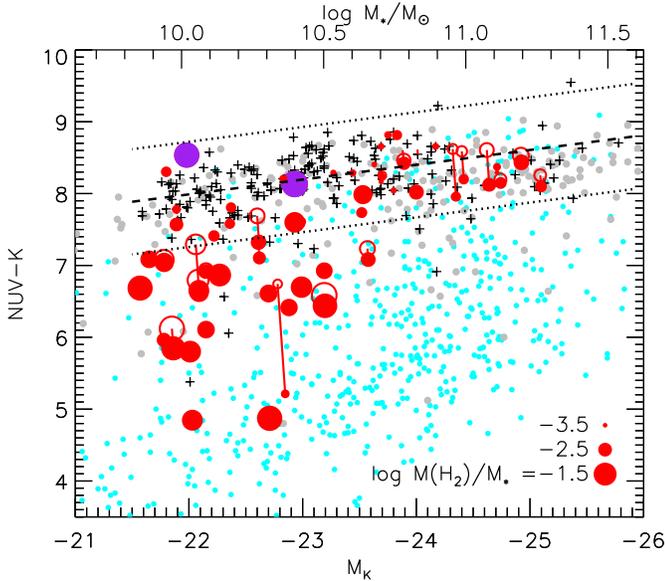}
\caption{Similar to Figure \ref{fig:galex_cmd}, except that galaxies
with central \hi\ detections have been added, and corrections for internal 
extinction have been made for the galaxies in Table \ref{tab:ar}.  Corrected
photometry values (filled circles) are linked with a line to the values before
correction (open circles).
\label{fig:galex_cmd_corr}
}
\end{figure}

Figures \ref{fig:CMD_corr} and \ref{fig:galex_cmd_corr} are analogous to Figures
\ref{fig:CMD} and \ref{fig:galex_cmd}, but the photometry has been corrected for
internal extinction in those cases where such correction is available.  
This correction makes only modest alterations to the colour-magnitude diagrams.
For example, in $u-r$, we find only four \htoo-rich 
galaxies that are on the red sequence before correction and off the red sequence 
after correction (though, as mentioned above, NGC\,1266, UGC\,09519, NGC\,4710, and
NGC\,5866 might also move off the red sequence with an internal extinction
correction.)
The highest mass, \htoo-rich red sequence galaxies remain on the red
sequence even with the extinction correction.  
We note also that the red sequence ridgelines and $2\sigma$ boundaries (Section
\ref{sec:implications}) were defined using only CO-nondetected galaxies, so for practical
purposes they are not affected by the internal extinction correction.

\section{Evidence for Gas accretion}\label{sec:accretion}

One possible interpretation of the cold gas detection rates in Table \ref{tab:redseq} is
that early-type galaxies occasionally retain some 
atomic and/or molecular gas throughout their
complex evolutionary histories, and through their transformations from late-type
to early-type.  
Another is that they may have lost their original cold gas but acquired some new
cold gas in recent times.
As discussed by \citet{davis11} and \citet{mcd13},
the kinematic and spectroscopic data of \atlas\ identify
specific galaxies that have experienced recent gas accretion.
The locations of these galaxies in the colour-magnitude diagrams can help to clarify 
which of the alternatives mentioned above may be more relevant to the development
of the red sequence and the histories of the `blue tail' early-type galaxies.

\subsection{Identifying signatures of cold gas accretion or interactions}

Kinematic misalignments between cold gas and stars indicate either that the gas was
accreted, and its incoming orbital angular momentum was not parallel to
the galaxy's angular momentum, or that the gas endured through some major
disruption such as a merger.  Thus the misalignments serve as markers  
of a galaxy's history.
Kinematic misalignments can also occur if the galaxy's potential is significantly
triaxial or barred.  Triaxiality might be important for a few of the slow 
rotators in the sample, but it should not be important for most of the CO
detections; a strong majority of them occur in fast rotators (Paper IV), and the fast rotators 
are consistent with being oblate 
\citep{paper2,paper3,weijmans,paper20}.  

The kinematic misalignment angles between stars, molecular gas, and ionized gas
are provided by \citet{davis11}, who
further show that the molecular gas kinematic position angles
(PAs) are always consistent with ionized gas kinematic PAs where both can be
measured.  Similarly, misalignment angles between central \hi\ and stellar
kinematics are provided by \citet{paolo-kin}.
Thus, galaxies are classified as having aligned gas or misaligned gas, with a 
dividing line at 30$^\circ$ which makes generous allowance for the uncertainties in
PA measurement and for the presence of bars.
Here we assume that all misaligned
gas was accreted from some external source.
This assumption will provide only a lower limit on the incidence of accreted gas, of
course, as some of the aligned gas may also have been accreted.

\citet{mcd13} have also identified a subset of \atlas\ galaxies which exhibit
unusually low stellar metallicities and strong $\alpha$-element enhancement, when compared
to their peers at similar stellar masses.  The low metallicity/strong [$\alpha$/Fe]
stellar populations are concentrated in the centers of their host galaxies, and
the hosts tend to be the most \htoo-rich sample members in low-density
environments.  \citet{mcd13} propose that these galaxies have an underlying old
stellar population which follows the same metallicity and $\alpha$-element
abundance trends as the rest of the sample, but they have accreted a substantial
amount of low metallicity gas which has formed a younger and more metal-poor
stellar population.  The galaxies thus identified are
IC\,0676, IC\,1024, NGC\,2764, NGC\,3073, NGC\,4684, NGC\,7465, PGC\,056772, PGC\,058114,
PGC\,061468, and UGC\,05408, and in the following sections they are also treated as
showing signs of gas accretion.

Another indicator of a galaxy's interaction and/or gas accretion history can 
come from disturbances to the gas distribution and kinematics.
We make qualitative assessment of such disturbances from the CO position-velocity
slices in \citet{davis13} and from the CO integrated intensity maps and velocity
fields in \citet{DRpaper}.
Markers of disturbances include gas tails, 
strong asymmetries in the gas distribution, and complex
velocity fields, and we thereby consider the following galaxies to have disturbed
molecular gas: IC\,1024, NGC\,1222, NGC\,3619, NGC\,4150, NGC\,4550,
NGC\,4694, NGC\,4753, NGC\,5173, NGC\,7465, PGC\,058114, and UGC\,09519.
These are the galaxies noted in \citet{DRpaper} as classes `M' (mildly disturbed)
or `X' (strongly disturbed), or which are sufficiently lopsided that their
integrated CO spectrum from the CARMA data has a flux density twice as large on one
side of the systemic velocity as on the other side.
NGC\,4550 is also noted by \citet{crocker4550} to have an asymmetric gas disc,
based on 
observations from the IRAM Plateau de Bure interferometer.
These classifications thus show disturbances in 11 of the
40 galaxies with CO maps.
Additional possible cases of disturbed gas include NGC\,2768 and
NGC\,3489 \citep{crocker-all}.
The qualitative assessments tend to be sensitive only to rather severe
disruptions, and furthermore the sensitivity is undoubtedly nonuniform, but the
classifications are adequate in these initial searches for large scale trends
between disruptions and galaxy colour.

Kinematic disturbances are also evident in the atomic gas of many \hi-rich
galaxies, especially at several $R_e$ and beyond
\citep{paolo,paolo-kin}.  
However, since that material at large radii
does not appear to be associated with recent star formation activity within $R_e$
(Section \ref{atomicgas}), we focus here on the kinematic information from the gas
on kpc scales.  The \atlas\ sample includes 
15 galaxies with central HI detections from WSRT, but no CO map; of these 15, 
only one (NGC\,7280) is identified as likely showing significant kinematically
disturbed gas.  We suspect that more candidates would be identified if the spatial
resolution of the HI data were closer to that of the CO data.

The disturbance indicator provides a different perspective
than the misalignment criterion discussed above, as it is useful
even in cases with well-aligned stellar and gaseous angular momenta. 
It is also most sensitive to events within the last few orbital time-scales
\citep[typically a few $\times 10^7$ to $10^8$ yr,][]{y02}, and this may be the most
appropriate time-scale for correlations with colour evolution.
However, it may indicate tidal forces on pre-existing gas in a galaxy, rather than
the accretion of new gas.
Furthermore, as it is clearly dependent on the sensitivity and
angular resolution of the images, we can at present only detect the most egregious
disturbances.
Deep optical images can help to distinguish between tidal interactions and gas
accretion \citep{megacam1}, and such work is ongoing, but for now we will take the
disturbance indicator as a possible sign of accretion with these caveats.

\subsection{Stellar populations vs.\ accretion/interaction signatures}

\begin{figure}
\includegraphics[scale=0.55,trim=1.5cm 0.5cm 0.5cm 1cm]{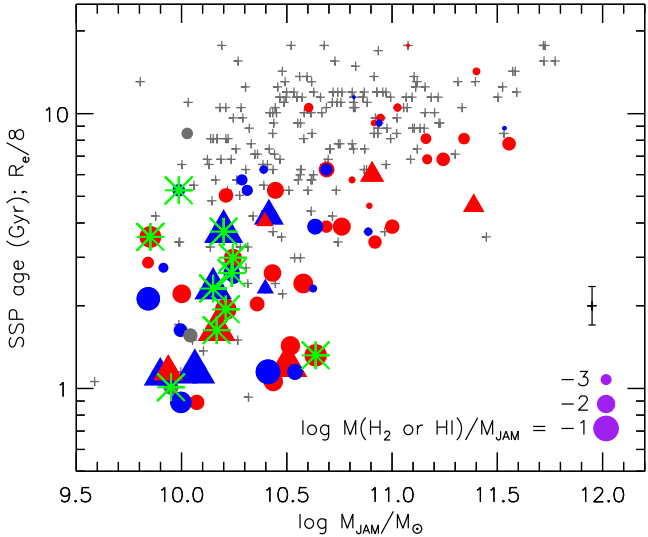}
\caption{Similar to Figure \ref{fig:agejam_cmd}, but here the circles and triangles
indicate galaxies with either CO or central \hi\ detections.  
Kinematically misaligned cold gas is indicated in blue and aligned (prograde) gas
is in red. 
Galaxies showing significant disturbances in the molecular gas or \hi\ (NGC\,7280) are indicated with
triangles and those showing relaxed gas are in circles.
In all cases the symbol size indicates
the scaled \htoo\ mass, if CO is detected, or the \hi\ mass if CO is not detected.
Outliers
in the stellar mass-metallicity and mass-[$\alpha$/Fe] relations are marked with
green stars.
The presence or absence of disturbances can only be noted for galaxies with CO or
\hi\
maps, and because only 70\% of the CO detections are mapped we have only lower
limits on the incidence of disturbances (triangles).  For galaxies without CO maps we have
used the kinematic misalignments measured in ionized gas or \hi\
\citep{davis11,paolo-kin}.
NGC\,0509, NGC\,3073, and NGC\,4283 are plotted in grey 
because they have insufficient kinematic information in all gas phases.
\label{fig:misalignedCO}
}
\end{figure}

Figure \ref{fig:misalignedCO} shows an age-mass
diagram again but with markers indicating the galaxies that have
signs of gas accretion or gravitational interactions.
There are significant overlaps between the indicators, so that many
galaxies have two or three of the signs.
The \atlas\ sample contains 158 galaxies with $\logmjam \le 10.7$,
and 45 of them ($0.28 \pm 0.03$) are detected in CO emission or in central \hi\
emission.  We find that
21 of those 45 ($0.47 \pm 0.07$) show misaligned gas, which we
interpret as meaning they have accreted their gas from some external source.
Further, 26 of them ($0.58 \pm 0.07$) show misaligned gas
and/or anomalously low stellar metallicities,
and 28 of them ($0.62 \pm 0.07$) show misaligned gas, low
metallicities, and/or kinematic disturbances in the cold gas at radii $< R_e$.
Thus, more than half of the low (stellar) mass, \htoo-rich or 
\hi-rich early-type galaxies
appear to have accreted substantial amounts of cold gas.
Considering the multiple markers of accretion and interaction makes a modest
increase in this rate over considering misalignment alone.
In contrast, only six of the 22 high stellar mass galaxies with central cold gas
($0.27 \pm 0.09$) show any such signs.  

From analysis of the stellar structures and kinematics,
\citet{paper7,paper20} have suggested 
that fast rotator early-type galaxies are former spiral galaxies which have been 
quenched through some bulge growth processes
\citep[see also][]{cheung12,cappellari-apjl}.
The fast rotators might conceivably retain some molecular gas through this 
transition.
However, spirals almost always have kinematically aligned gas and stars
\citep[][and references therein]{bureauchung}.
We might thus expect that in the absence of a major merger or external accretion, 
quenched former spiral galaxies 
will have primarily aligned gas (like the red symbols in 
Figure \ref{fig:misalignedCO}) rather than strongly misaligned gas (blue symbols).

In this context, we consider whether the early-type galaxies with aligned gas might
have had different evolutionary histories from those with misaligned gas.  We
restrict attention at this point to the low (stellar) mass galaxies, due to the
small incidence of misaligned gas among massive galaxies.
Statistical testing shows no compelling evidence that the low-mass,
\hi-rich and \htoo-rich galaxies with misaligned gas differ from their counterparts
with aligned gas. 
Specifically, Kolmogorov-Smirnov and Mann-Whitney U tests on the galaxy
colours, equivalent SSP ages, metallicities, $\alpha$-element
abundances, stellar velocity dispersions, specific stellar angular momenta, and
photometric disc-to-total light
ratios give probabilities 0.13 to 0.99, none of which are small enough to infer a
difference in the two populations.  
Ultimately the question may be decided through detailed exploration of the
individual galaxies' star formation histories \citep{mcd13}, with deep optical
imaging and measurements of the
current star formation rates and gas depletion time-scales.
At present we conclude that the cold gas in early-type galaxies includes some gas
which may have been retained through the quenching transition to the red sequence,
as well as some gas acquired in a major merger or accreted from an external source
such as a satellite galaxy or the intergalactic medium.
This suggestion is 
not new, of course \citep[e.g.][]{thilker10}.  But for the first
time we now bring statistical samples of cold gas kinematic data to compare with
the stellar population and structural data.

\citet{kaviraj12} have advocated the idea that all the dust and molecular gas in 
early-type galaxies is accreted from external sources, based on an analysis of dust lane
galaxies in the Galaxy Zoo project.
It is worth noting, however, that
part of their argument is based on their sample's large inferred dust and
molecular masses,
\mhtoomstar $\sim$ 0.013.  The dust detection rate in
\citet{kaviraj12} is also substantially lower than our detection rate in CO (4\%
vs 22\%).  The dust lane galaxies identified in
\citet{kaviraj12} thus represent only the extreme tail of \htoo-rich early-type
galaxies, and the current study probes further into the \mhtoomstar\ luminosity
distribution.

Finally, in the context of reading the galaxies' histories
it is also important to note that the kinematic disturbances we identify
in the molecular gas may not be the result of accretion from some external source
onto a gas-poor galaxy.  Instead, they may be the result of a strong 
gravitational interaction on an already
gas-rich early-type galaxy, or (in the case of lopsidedness) an internal
instability.  To test for these different possibilities,
Table \ref{tab:classes} gives the incidence of disturbances in the CO kinematics
for the low-mass galaxies with both aligned and misaligned gas.
We note that disturbances are somewhat more common among
low-mass galaxies with misaligned gas $(5/12 = 0.42 \pm 0.14)$ than among those with aligned
gas $(6/27 = 0.22 \pm 0.08)$. However, the Fisher exact and
chi-squared contingency table tests do not give better than 95\% confidence levels
on rejecting the null hypothesis, so the sample sizes are still too small to
associate disturbances preferentially with gas accretion (as opposed to outflow or
tidal forces).

\begin{table}
\caption{Misalignments and Disturbances in low-mass \htoo-rich galaxies\label{tab:classes}}
\begin{tabular}{lccc}
\hline
{} & {Relaxed CO} & {Disturbed CO} & {Total} \\ 
\hline
aligned & 21 & 6 & 27 \\
misaligned & 7 & 5 & 12 \\
Total  &  28 & 11 & 39 \\
\hline
\end{tabular}
\medskip

\textit{Notes:} UGC\,05408 has such a poorly resolved CO distribution that it is
not possible to classify as either relaxed or disturbed.
\end{table}

\subsection{Models of accretion-driven colour and \hbeta\ evolution}

Extensive evidence for the accretion of cold gas, 
particularly in the low-mass `blue
tail' galaxies, raises additional questions about the colour evolution of early-type
galaxies.
For example, we may ask whether it is possible for a completely quiescent red
sequence galaxy to become as blue as our observed blue galaxies, given the
constraints of the observed molecular masses and star formation rates. 
Secondly, we may
ask whether cold gas accretion onto a massive galaxy, with an associated burst of star
formation, would necessarily drive the galaxy off the red sequence.

\begin{figure}
\includegraphics[scale=0.55,trim=1.5cm 0.5cm 0.5cm 1cm]{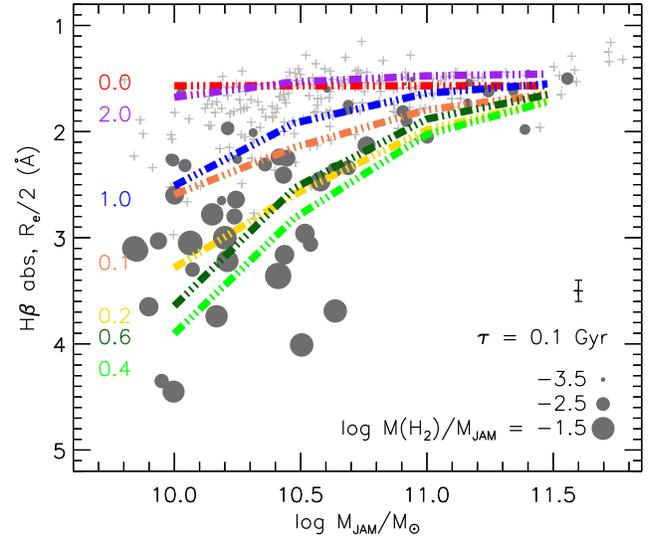}
\caption{Similar to Figure \ref{fig:hbeta_cmd}, with overlays showing
the evolution of old galaxies with 3\e{8} \solmass\ of recent star formation activity.  The 
red line marked `0.0' shows the model red sequence before star formation is initiated,
and subsequent lines show how \hbeta\ changes through the late burst.  Each
isochrone is labelled with the time (in Gyr) since the initiation of the burst.
\label{fig:toy1}
}
\end{figure}

Here we explore these questions by constructing
a set of toy models in which quiescent, red sequence galaxies
of masses $\logmstar = 10$ to 11.5
experience 3\e{8} \solmass\ of star formation activity at late times.
Detailed information on the models is provided in the Appendix; 
briefly, we use the Flexible Stellar Population Synthesis codes of \citet{conroy} to
follow their colour and \hbeta\ absorption line strength evolution through the star
formation episode.
The duration and strength of the `blue tail' excursion is, of course, 
determined by the assumed mass of young stars and the time-scale over which the late star formation occurs. 
The mass of young stars in most \atlas\ galaxies is poorly constrained, but many of the
blue tail galaxies have current molecular masses greater than 3\e{8}~\solmass,
which at least lends plausibility to our assumption.
In Fig \ref{fig:toy1}
we show one example that follows the evolution of the \hbeta\ absorption line 
strength through a burst of $\dot{M} \propto t
\exp(-t/\tau)$, with $\tau = 0.1$ Gyr.  
The burst begins $(t=0)$ about 10 Gyr after most of the galaxy's stars formed,
and the star formation rate
peaks at a time $t=\tau$ and subsequently decays on a time-scale of $\tau$.
The \hbeta\ line strength increases to a maximum of about 4\AA\ for the lowest-mass
galaxies and then, over a period of a couple Gyr, returns to its pre-burst value.
In a similar spirit, \citet{scott12} have shown that \atlas\ galaxies with young
central populations and low central Mg~$b$ values will return to the nominal
Mg~$b$--$\mathrm{V_{esc}}$ relation after several Gyr of passive evolution.
The behaviors for other time-scales and for $u-r$ and \nuvk\ colours are also shown 
in the
Appendix.

The toy models clearly show that it is plausible to explain the blue tail galaxies
as former red sequence galaxies, with a quantity of recent star formation which is
consistent with the observed \htoo\ masses in the \atlas\ sample.  The required
star formation rates are also
consistent with those observed \citep[][Appendix A]{crocker-all,shapiro}.
In this excursion scenario relatively short star formation time-scales, $\tau \sim 0.1$  to 0.3 Gyr, 
are required in order to reproduce the bluest galaxies with the strongest \hbeta\
absorption.
For the massive galaxies, a tight red sequence (particularly in \nuvk) favors
longer star formation time-scales or older bursts, as discussed in more detail in
the Appendix.
Detailed studies of the galaxies' star formation histories are necessary for moving
beyond this very simple plausibility check, but the models are consistent with a
picture in which gas accretion onto red sequence galaxies could produce at least 
some of the blue tail galaxies.

Galaxy mergers are often associated with enhanced star formation activity, as
in the case of the luminous and ultraluminous IR galaxies (LIRGs and ULIRGs); the
mechanism is thought to be the galaxy-scale shocks driving the molecular gas to
higher densities.  In this context it is interesting to pursue the question
of whether the galaxies with kinematic signs of gas accretion or disturbance have
shorter star formation time-scales than the galaxies with quiescent kinematics.  
As noted above, such differences are not measurable in integrated colours or the $R_e/2$
\hbeta\ absorption strengths.
\citet{davis-sf} also find that such differences are not obvious in star formation
rates measured from mid-IR emission.

Some recent studies of the galaxy colour-magnitude diagram \citep[e.g.][]{goncalves}
have assumed that galaxies in the `green valley' are exclusively
becoming redder, on the approach to the red sequence.  At face value, such an 
assumption would seem to be in direct contradiction of the toy model employed here.
However, our toy model does apply to a somewhat special population of galaxies, the
\htoo-rich early-type galaxies.  In addition, their movement through the green
valley is faster on the blueward track than on the return redward track, so even
in the context of this `excursion' model there will be a greater proportion of
galaxies moving redward than moving blueward.

\section{High-mass vs.\ low-mass galaxies}\label{sec:highmass}

Section \ref{sec:accretion} discusses evidence that the cold gas in low-mass early-type
galaxies is commonly accreted.  However, the picture is much less clear
for the high-mass \htoo-rich
galaxies, since their incidence of disturbed or misaligned gas is lower. 
The detectability of kinematic misalignments
and other gas disturbances should not depend on the 
stellar mass of the galaxy,
except to the extent that the disturbed cold gas will evolve on an orbital
time-scale 
towards circular orbits in a stable plane.  In our mostly oblate
gas-rich galaxies, the
evolution will be towards prograde or retrograde orbits in the equatorial plane,
and the orbital time-scales are
shorter for a larger galaxy mass (at fixed radius).
Thus the difference in the rates of misalignment or gas disturbance between
low-mass and high-mass galaxies is probably real, though  
it may simply mean that the high-mass galaxies accreted their gas in the more
distant past. For further discussion, see \citet{davis11} and \citet{sarzi06}.
On the other hand, the metallicity signatures of
accreted gas should be more easily distinguished in a smaller host stellar population.
The catalog of low-metallicity accreted gas is expected to be
strongly incomplete for large stellar masses.

In a study of the hot gas in \atlas\ galaxies, \citet{sarzi} note that
interactions between hot and cold gas may have a significant effect on the cold
gas.  They propose that the higher mass fast rotators, due to their deeper
gravitational potentials, may have recycled more of their internal stellar mass
loss into a hot ISM.  
This recycled gas would necessarily be kinematically
aligned with the stellar rotation, and it might shock heat any incoming
misaligned gas (thus producing the lack of misaligned molecular gas in high-mass
galaxies).   \citet{bogdan12} have also presented a case study of the
hot gas content of four high-luminosity $(M_K \sim -24.0)$ \atlas\ galaxies, and
they demonstrate that early-type galaxies of similar optical luminosity can have 
hot gas of very different distributions and kinematic states.  A proper treatment
of the interaction of hot gas and cold gas should take into account the possible
angular momentum transfer between the gas phases.

\begin{figure}
\includegraphics[scale=0.5,trim=0cm 0cm 0cm 0cm]{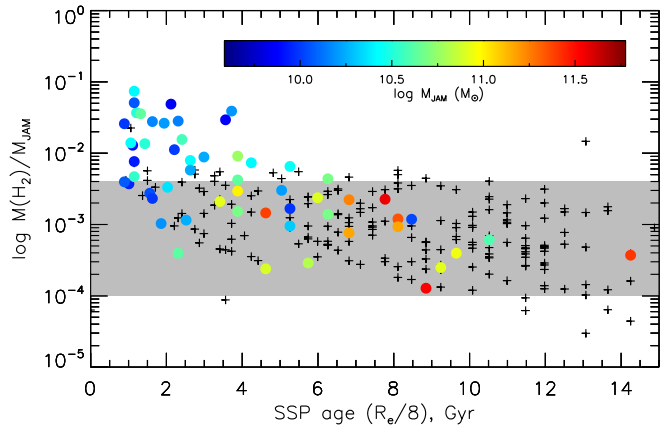}
\caption{Molecular gas content vs.\ SSP age; crosses are
CO nondetections, and colour circles are CO detections.  The colour of each 
circle indicates the
total stellar mass $\log(M_{\text{JAM}}/M_\odot)$ as noted in the colourbar.
As expected, there is a general trend such that galaxies with larger molecular
contents tend to have younger stellar ages.  In addition,
for \mhtoomjam\ in the range
$10^{-2.4}$ to $10^{-4}$ (gray box), galaxies of larger stellar mass \mjam\ have systematically
larger SSP ages even though they have the same molecular gas content (normalized to
stellar mass). 
\label{fig:mh2hbcolor}
}
\end{figure}

We have commented several times that there is a striking difference in the SSP age
and \hbeta\ absorption line strength scatter of high and low-mass early-type galaxies.  
\citet{scott12} have also made the same comment about the
Mg~$b$ line strengths and escape velocities in the \atlas\ sample.
In colour and \hbeta, some of this
difference is due to the relative gas masses; specifically, the high-mass 
galaxies tend to have smaller \mhtoomjam\ ratios.
However, as Figure \ref{fig:mh2hbcolor} shows, there is a significant range of overlap
and values of \mhtoomstar\ from $10^{-2.4}$ to $10^{-4}$ can be found in both high-mass
and low-mass early-type galaxies.  Under the very simplistic assumption that a fixed
\mhtoomstar\ corresponds to a fixed percentage of young stars in a late starburst, 
one might then expect all galaxies of the same \mhtoomstar\ to have the same SSP age.
Instead, we observe that high-mass galaxies have
systematically larger SSP ages even at the {\it same}
values of \mhtoomstar\ or \mhtoomjam\ as low-mass galaxies.  
The trend suggests differences in the star formation efficiencies and/or the
accretion and star formation histories between low-mass and high-mass galaxies.
\citet{naab} discuss the star formation and assembly histories of
simulated high-mass and low-mass early-type galaxies, for example, and note that the
high-mass galaxies have older stellar populations.
The differences might also be related to the fact that the high-mass galaxies tend to have
kinematically quiescent, prograde gas, so presumably they have fewer internal shocks
to drive enhanced star formation.

In addition, the morphological quenching
mechanism studied by \citet{kawata} and \citet{martig12} (among others)
might explain this trend, as it invokes greater local stability of the gas 
in a deeper and steeper gravitational potential.
For example, \citet{martig12} make high-resolution simulations of a cold gas disc
$(7.5\times 10^8~ \rm M_\odot)$ in a galaxy with $\rm M_\star = 5.6\times 10^{10}~\rm
M_\odot$; this galaxy then has \mhtoomstar $\sim 0.013,$ similar to the most
\htoo-rich of our high-mass early-type galaxies.  If the galaxy's stellar mass 
distribution
is spherical, they find the cold gas only reaches maximum densities of 600\percc,
whereas if the stellar distribution is flattened like a spiral, the cold gas reaches
4\e{5}\percc.  This difference suggests the same gas will have substantially longer
depletion time-scales in a spherical galaxy than in a flat galaxy. 
For specific application to the \atlas\ sample, we note that
\citet{paper20} have demonstrated that the high-mass early-type galaxies tend to be
spheroid-dominated whereas the low-mass galaxies include both spheroid-dominated
and disc-dominated systems.
One remaining question is whether
this kind of a quenching mechanism by itself can explain the apparent sharp and
dramatic increase in colour dispersion at low masses or whether, as mentioned above,
some systematic difference in gas accretion histories is also required.
These issues can be addressed in more detail by comparing the
star formation histories and current gas depletion time-scales of \atlas\
galaxies.

There are several other differences between the dust and ISM properties of
high-mass and low-mass early-type galaxies.
\citet{dsa12} find that higher mass Virgo Cluster early types tend to have higher
dust temperatures than their counterparts at lower stellar mass.
The reason for the trend is not yet known.
\citet{crocker-HD} have also noted correlations between stellar mass and the 
physical properties of the molecular gas in \atlas\ galaxies.  The high-mass
galaxies tend to have larger $^{13}$CO/$^{12}$CO ratios and larger 
HCN/$^{12}$CO ratios.  
The proposed interpretation of these trends is that they are driven by the
kinematic state of the gas (as noted above, the high-mass galaxies are less likely
to have disturbed or misaligned molecular gas).
But the physical properties of the molecular gas both affect, and are
affected by, star formation activity, and with the existing data it is still
difficult to disentangle these causal relationships from environmental effects
such as stripping in clusters and groups.  Additional work on the details of the
molecular gas properties would be valuable for understanding the star formation
efficiencies of early-type galaxies.

\section{SUMMARY}
\label{sec:summary}

We present a study of the relationships between the cold gas (\hi\ and \htoo)
content of early-type galaxies and their stellar populations, as encoded in
\hbeta\ absorption line strengths, $u-r$ and \nuvk\ colours.
Detection rates of \hi\ and \htoo\ are $\ge 50\%$ in `blue tail' early-type galaxies,
depending somewhat on how the blue tail is defined. 
Similarly, the low-mass 
\htoo-rich early-type galaxies are usually more blue and have younger stellar
populations than their \htoo-poor
counterparts,
especially when measured in \hbeta\ absorption and \nuvk\ colours, and less frequently at $u-r$.

Detection rates of \hi\ and \htoo\ are also nonzero for red sequence early-type 
galaxies, {\it no matter how the red sequence is defined}. 
These detection rates range from 10\% to 30\%
for galaxies with masses $\logmjam = 9.8$ to 11.8 and for different definitons of
the red sequence.
The result is at least
qualitatively consistent with the detection rates
of dust in red sequence early-type galaxies \citep[e.g.][]{smith12}.
The red sequence thus contains high mass early-type galaxies with abundant
cold gas; specifically, with
\htoo\ masses up to 8\e{8} \solmass\ and \hi\
masses up to 4\e{9} \solmass.
These results may serve as a reminder that `photometric gas
fractions' are merely averages over the relevant populations, and individual
galaxies can deviate strongly from those averages.
Colour corrections for the reddening associated with the molecular gas and dust do
not materially alter these conclusions.

The \htoo-rich and \hi-rich massive early-type galaxies have the same median colour
and same colour dispersion as their analogs without cold gas. 
To first order, the reason for this behavior is straightforward.
The massive galaxies tend to have smaller $\rm M_{gas}/\rm M_\star$, so they probably
also have smaller ratios $\rm M_{new}/M_{old}$ (where $\rm M_{new}$ is the mass of young
stars and $\rm M_{old}$ is the mass of old stars).  The young stellar populations will
therefore be more difficult to detect via their colour effects on a large host galaxy.

However, $\rm M_{gas}/\rm M_\star$, or even \mhtoomstar, does not uniquely predict the colour and mean
stellar age of an early-type galaxy; 
we note that massive galaxies have older stellar populations 
than low-mass galaxies {\it even at the same values of}
$\rm M_{gas}/\rm M_\star$.  We speculate that the low-mass galaxies may have shorter
gas depletion time-scales or that they may have acquired their cold gas more
recently.  Additional high sensitivity, high resolution molecular and radio
continuum observations (such as are now available with ALMA and the JVLA) will
complement improved simulations \citep[e.g.][]{martig12} in addressing these questions.

Atomic gas in early-type galaxies is not as closely related to star formation
activity as molecular gas is.  Many \hi-rich early-type galaxies are
still on the red sequence, at all stellar masses and in all indicators of colour,
\hbeta\ absorption and SSP age.
That result is largely driven by the differing spatial scales of the 
measurements.  Specifically,
the stellar population indicators we are working with here are dominated by
stars within an effective radius or less, but much of the \hi\ is at several $R_e$ and beyond
\citep{paolo}.
At the moment we are not considering young stellar populations in extended UV rings.
Furthermore, {\it if} the large-scale \hi\ distributions were serving as reservoirs
of cold gas to drive star formation activity in the cores of their galaxies, we
might expect total \hi\ content to correlate with colour, \hbeta\ strength or SSP age
 in the same sense that the \htoo\ mass does.
But this correlation is not observed in the extended \hi, only in the
central \hi.  
The time-scales over which the extended \hi\ can serve as a reservoir may be too long for us to
notice the correlation; or we might be
finding these extended \hi\ systems
in a stage prior to conversion of \hi\ into \htoo\ and star formation
activity. 
 
These relationships between cold gas content, galaxy colour, and \hbeta\ absorption
can produce selection biases in samples of red sequence galaxies.
For example, 
the detection rates of cold gas are larger among
the massive red sequence galaxies than the low-mass red sequence galaxies.

We have also analyzed the stellar properties of galaxies with and without signs
that their cold gas was accreted from some external source.
These signs include kinematic misalignments between gas and stars, disturbed gas
kinematics, and departures from the stellar mass-metallicity relation.
As \citet{davis11} have already commented, kinematic misalignments are not
common at high stellar masses or among the Virgo Cluster members, 
so we infer that $\sim 20\%$ of the high-mass early-type galaxies
galaxies may have retained or recycled their cold interstellar medium.
However, at low stellar masses, $\sim 60\%$
of the \hi- and \htoo-rich galaxies show at least one of the signs that the 
cold gas was accreted. 

Simple toy models can explain the colours and \hbeta\ absorption strengths of the
\htoo-rich \atlas\ galaxies in a ``frosting" or rejuvenation scenario.  Even the
bluest of them
could be formerly quiescent, red sequence galaxies which acquired
a modest amount of cold gas and recent star formation.  
The required star formation rates and gas masses are broadly consistent with 
those observed.  Of course, this colour analysis provides only a plausibility
argument for the rejuvenation scenario and more detailed work on the galaxies' star
formation histories is necessary.

We have searched for
significant structural or stellar population differences between the gas-rich
galaxies that clearly show signs of external accretion, and the gas-rich galaxies
with relaxed, kinematically aligned gas.  No
notable differences are found.  But some of the prograde, well-aligned cold gas may
also have been accreted, and
additional theoretical work on the expected angular momentum distributions of
accreted material would be valuable.

Some of the implications of this work include the idea that
red sequence galaxies should not be assumed to be free of cold gas, even
if a morphological cut is made to exclude red spirals and retain only red early-type
galaxies.  Early-type galaxies in the green valley may have
recently acquired cold gas, and may still be becoming bluer.
Furthermore, for galaxies of $\logmjam > 10.7$ and
cold gas masses up to a few $10^9$ \solmass, and for the sample size we have
here,
the presence of \htoo\ and \hi\ do not measurably affect
integrated colours, the red sequence ridgeline or its colour dispersion.  
Hence, within these parameters, great care must be taken to
make inferences about the cold gas content and star formation rates of galaxies 
from the colour evolution of the red sequence or the lack of such evolution.

\section*{Acknowledgments}
This research was partially supported by grant NSF-1109803 to LMY.
Thanks also to Dr.\ Paul T.~P.~Ho for the invitation to spend a sabbatical at
ASIAA.
MC acknowledges support from a Royal Society University Research Fellowship.
This work was supported by the rolling grants ‘Astrophysics at Oxford’ PP/E001114/1
and ST/H002456/1 and visitors grants PPA/V/S/2002/00553, PP/E001564/1 and
ST/H504862/1 from the UK Research Councils. RLD acknowledges travel and computer
grants from Christ Church, Oxford and support from the Royal Society in the form of
a Wolfson Merit Award 502011.K502/jd. RLD is also grateful for support from the
Australian Astronomical Observatory Distinguished Visitors programme, the ARC
Centre of Excellence for All Sky Astrophysics, and the University of Sydney during
a sabbatical visit.
SK acknowledges support from the Royal Society Joint Projects Grant JP0869822.
RMcD is supported by the Gemini Observatory, which is operated by the Association
of Universities for Research in Astronomy, Inc., on behalf of the international
Gemini partnership of Argentina, Australia, Brazil, Canada, Chile, the United
Kingdom, and the United States of America.
TN and MBois acknowledge support from the DFG Cluster of Excellence `Origin and
Structure of the Universe'.
MS acknowledges support from a STFC Advanced Fellowship ST/F009186/1.
PS acknowledges support of a NWO/Veni grant.
(TAD) The research leading to these results has received funding from the European
Community's Seventh Framework Programme (/FP7/2007-2013/) under grant agreement
No 229517.
MBois has received, during this research, funding from the European Research
Council under the Advanced Grant Program Num 267399-Momentum.
The authors acknowledge financial support from ESO. 

Funding for SDSS-III has been provided by the Alfred P. Sloan Foundation, the
Participating Institutions, the National Science Foundation, and the U.S.
Department of Energy Office of Science. The SDSS-III web site is
http://www.sdss3.org/.

SDSS-III is managed by the Astrophysical Research Consortium for the Participating
Institutions of the SDSS-III Collaboration including the University of Arizona,
the Brazilian Participation Group, Brookhaven National Laboratory, University of
Cambridge, Carnegie Mellon University, University of Florida, the French
Participation Group, the German Participation Group, Harvard University, the
Instituto de Astrofisica de Canarias, the Michigan State/Notre Dame/JINA
Participation Group, Johns Hopkins University, Lawrence Berkeley National
Laboratory, Max Planck Institute for Astrophysics, Max Planck Institute for
Extraterrestrial Physics, New Mexico State University, New York University, Ohio
State University, Pennsylvania State University, University of Portsmouth,
Princeton University, the Spanish Participation Group, University of Tokyo,
University of Utah, Vanderbilt University, University of Virginia, University of
Washington, and Yale University.

\appendix
\section{Toy Models for Colour and \hbeta\ evolution\label{sec:toymodels}}

We make a quantitative assessment of the effects of recent star formation on red
sequence galaxies using the Flexible Stellar Population Synthesis codes (version
2.4) of
\citet{conroy}.  Specifically, we model a scenario in which a relatively small
amount of cold molecular gas drops into a red sequence galaxy and ignites star
formation \citep{ch09,kaviraj11},
and we follow the colour and \hbeta\ evolution of these simple models.
We use the Padova isochrones and a Chabrier initial mass function.  For following
integrated colours, the BaSeL spectral library is used because it has a larger wavelength
coverage than MILES; but for 
\hbeta\ absorption, the MILES spectral library is used as it has the best match to the \atlas\
optical spectroscopic resolution.
The evolution is traced by first constructing models of old, red sequence galaxies
with ages of 8.5 to 10.3 Gyr, 
metallicities [Fe/H] of solar to +0.2 dex, and stellar masses appropriate to span the
range of luminosities in the \atlas\ sample.  The ages and metallicities of this
old population are adjusted
by hand to reproduce the slope of the red sequence, but that is purely for
cosmetic purposes and we
do not attempt to infer anything about the details of the old stellar
populations in this analysis.  
The \hbeta\ ``red sequence" is a simpler solar metallicity model with an age of
10.2 Gyr.
Late bursts of star formation are then superposed,
and the evolution of the composite population is followed to produce ``red
sequence" isochrones
following the burst.
No extinction is assumed on the young population, so the colour isochrones 
illustrate the limiting case of the maximum excursion from the red sequence.

The total mass of stars formed in the late burst is always $\int\dot{M} dt = 3\times
10^8$~\solmass, 
motivated by the fact that the largest molecular masses inferred
in the \atlas\ sample are on the order of 3\e{9}~\solmass.  
A stellar mass of 3\e{8}~\solmass\ is therefore reasonable in the sense that it is
smaller than many of the currently known molecular masses.  
Interestingly, an HST study of NGC\,4150 (one of our low-mass blue
tail galaxies) infers that it contains 1--2\e{8}~\solmass\ of young stars
\citep{crockett-4150}, though its current molecular mass is only 7\e{7}~\solmass\
\citep{paper4}.  In short, the mass of young stars can also be larger than the
current molecular mass if a significant portion of that gas has already been
consumed or destroyed.
The use of a constant
mass of young stars, independent of total stellar luminosity, is also motivated by
the statistical work in Paper IV showing that the \mhtoo\ distributions are
independent of stellar luminosity.  

Four sets of models
are produced for different star formation histories in the late burst; one set
assumes an instantaneous burst, and three are delayed $\tau$ models with $\dot{M}(t)
\propto t e^{-t/\tau}$ for $\tau = 100$~Myr, 300 Myr, and 1 Gyr.  
In this assumed form, the star formation rate peaks
at time $t=\tau$, so that $\tau$ describes the
time-scale for the initial ramp-up of the star formation rate from zero to its peak
as well as for the decay from the peak back down to zero.
The instantaneous burst ($\tau = 0$) is clearly unphysical; even in the case of an extreme
violent merger one would still expect gas to fall in on a dynamical time-scale, but
the model is presented for context.
The exponential cutoff in the star formation rate is imposed so that the models
will gradually return to the red sequence.  In the real universe such a cutoff
might be driven by consumption of the cold gas, by its destruction or expulsion in
AGN-driven feedback, or something else, but we are agnostic about the cause of the
cutoff and merely observe its effects on the stellar population.

The peak star formation
rates in the delayed $\tau$ models are 1.1 \solmass\peryr, 0.37 \solmass\peryr, and 0.11
\solmass\peryr.  In these models the star formation rate is within 90\% of
its peak value for the time period $0.6\tau$ to $1.5\tau$ and is significantly lower
outside that time range.
For comparison, the star formation rates 
inferred for the \atlas\ galaxies by \citet{shapiro}, 
\citet{crocker-all}, and \citet{davis-sf} from observations of PAH, 22\micron, 
24\micron\ 
and FUV emission are 0.006 to 3 \solmass\peryr\ and gas
consumption time-scales fall in the range 0.1 Gyr to 4 Gyr.
\citet{wei} also measured gas consumption time-scales in a different set of
early-type galaxies to be 0.3 to 10 Gyr.
Thus the delayed $\tau$ models are consistent with observed star formation rates.

\begin{figure*}
\includegraphics[scale=0.6,trim=2cm 0cm 1cm 0cm]{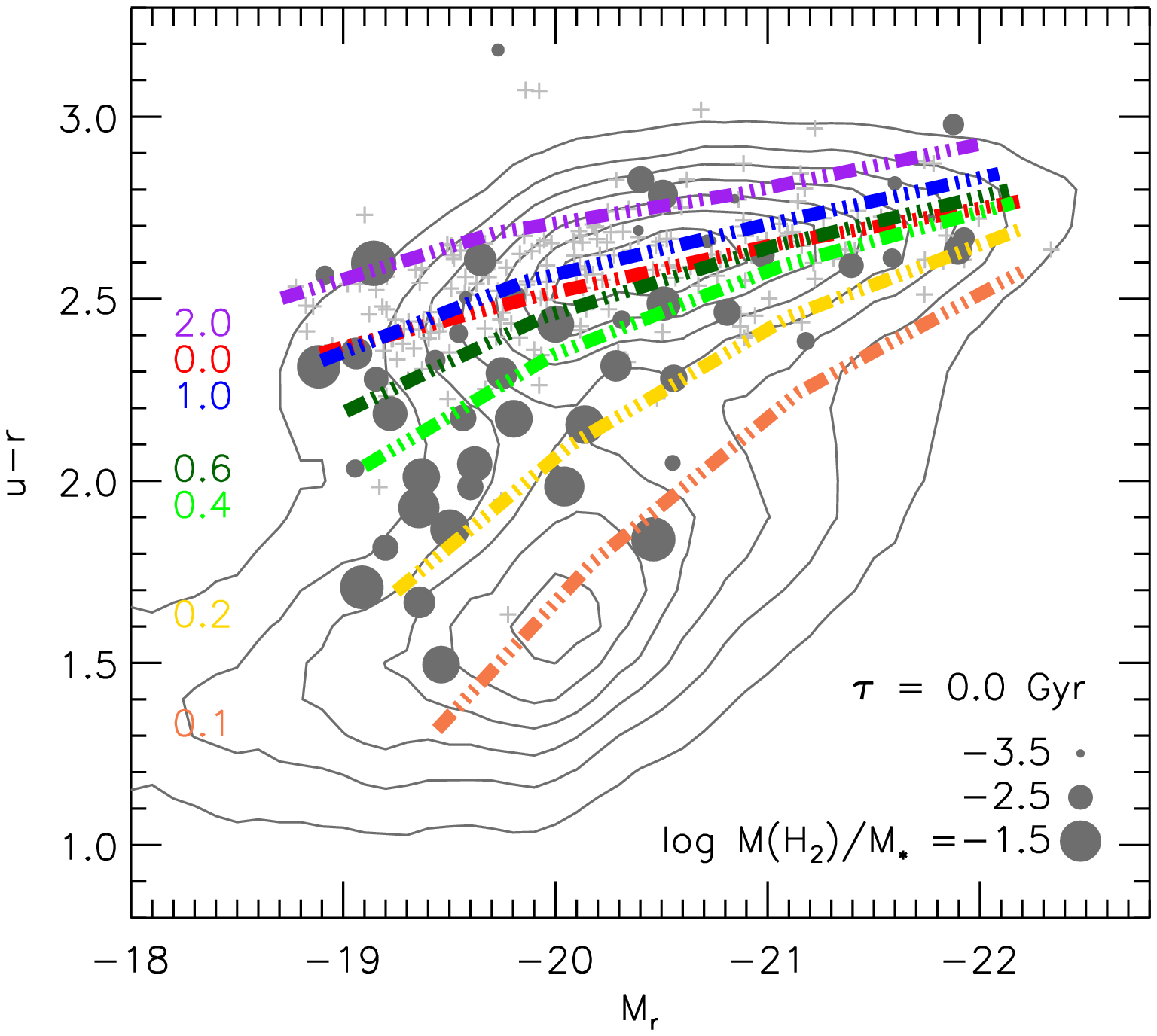}
\includegraphics[scale=0.6,clip,trim=3cm 0cm 0.8cm 0cm]{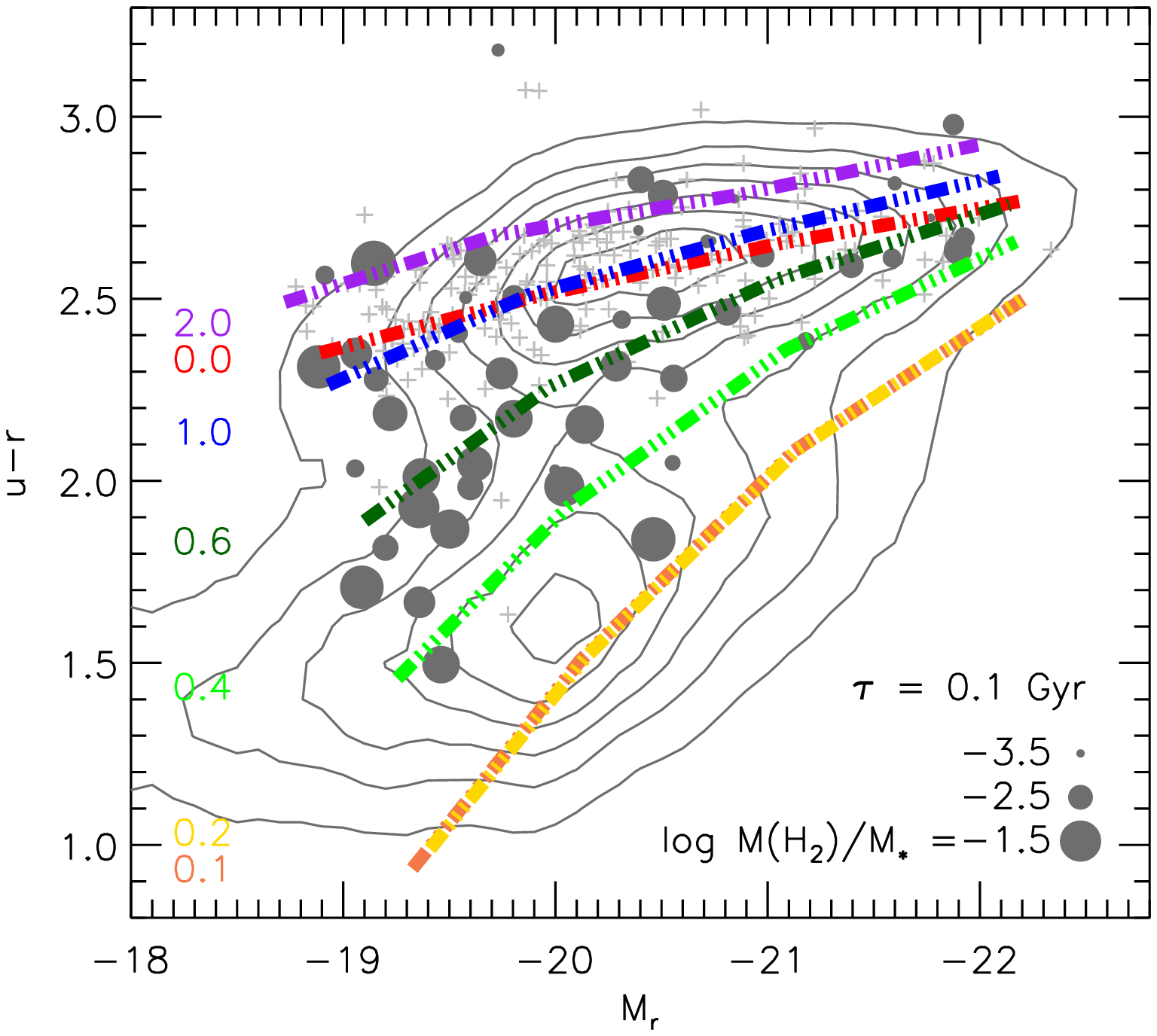}
\includegraphics[scale=0.6,trim=2cm 0cm 1cm 0cm]{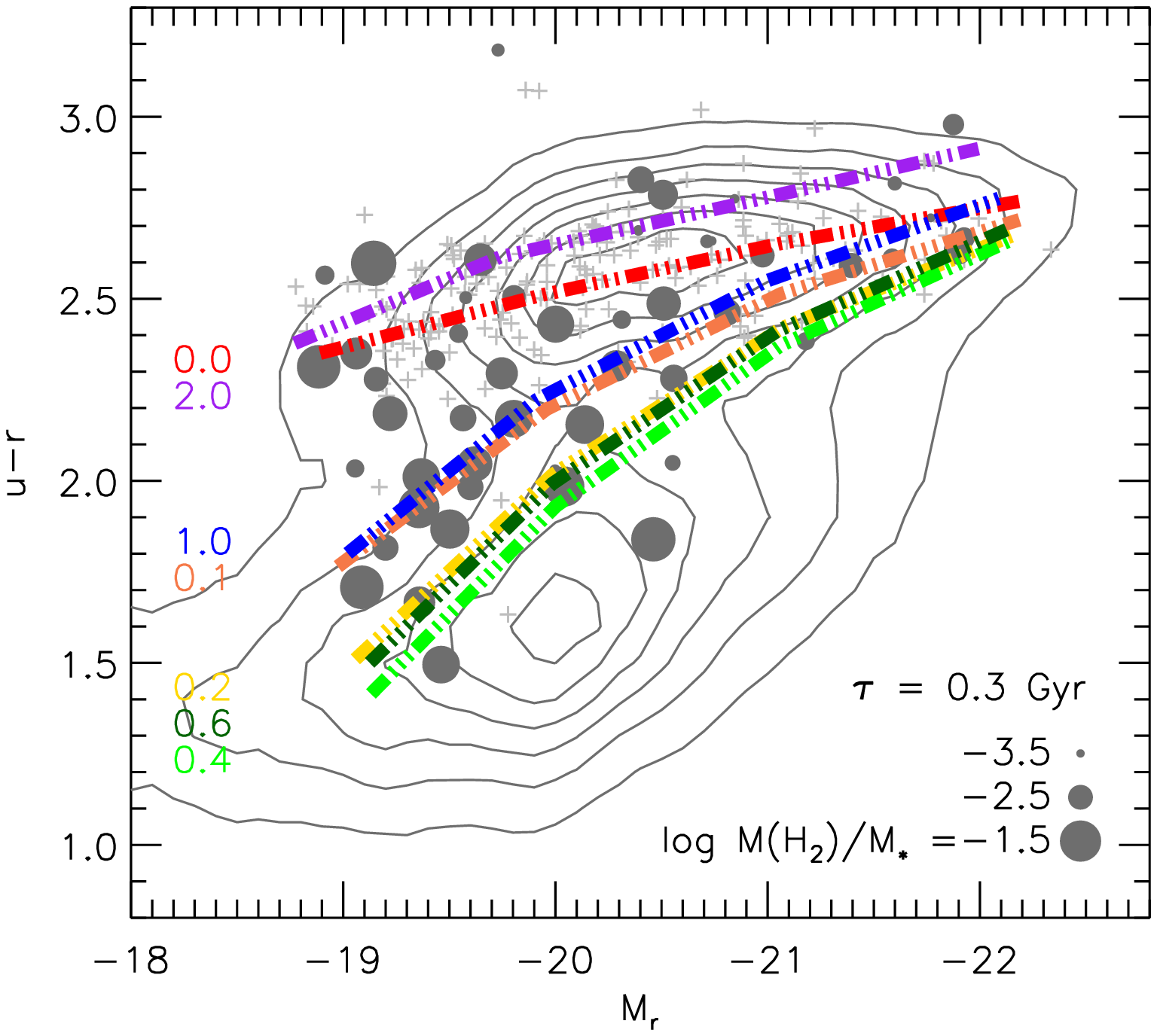}
\includegraphics[scale=0.6,clip,trim=3cm 0cm 0.8cm 0cm]{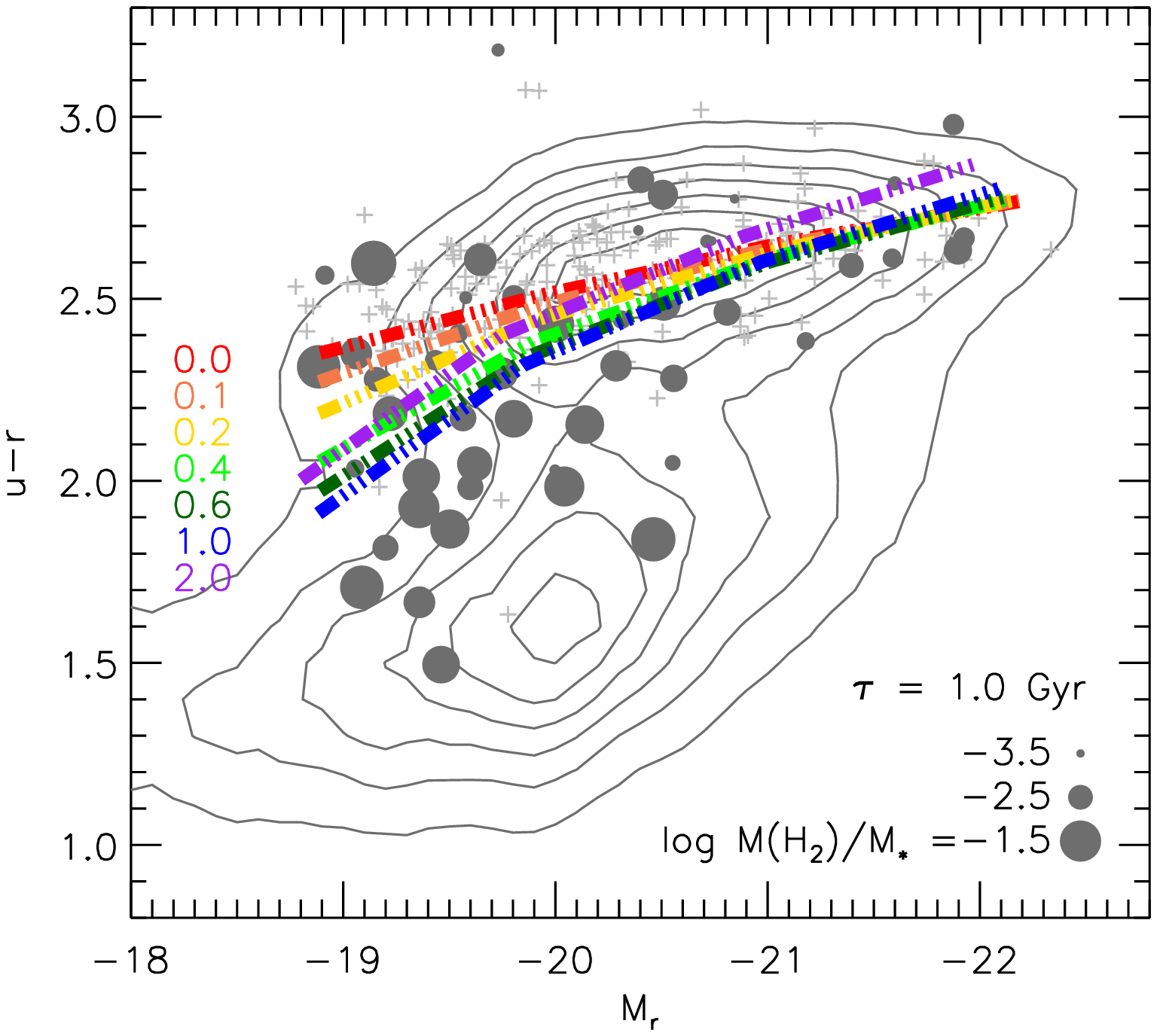}
\caption{Model isochrones for a simple scenario in which cold gas drops onto a red sequence
galaxy and initiates some star formation activity.  The total mass of stars formed is $3\times
10^8$ \solmass.  Panels show four different star formation histories for the late burst,
including a delta function ($\tau = 0$ Gyr) and delayed tau models with $\tau = 100$ Myr, 300
Myr, and
1 Gyr.  Dashed lines indicate the location of the ``red sequence" at the times
indicated (in Gyr, relative to the beginning of the burst).  Contours and other
symbols are the same as in Figure \ref{fig:CMD_corr}, with the exception that here the
symbol size is the normalized \htoo\ mass.
In the $\tau = 0$ panel, an isochrone earlier than 0.1 Gyr would show yet bluer
colours; in the other panels, the minimum in $u-r$ colours can be seen where the
isochrones change from becoming bluer with time to becoming redder.
\label{fig:toymodels}
}
\end{figure*}

\begin{figure*}
\includegraphics[scale=0.6,trim=2cm 0cm 1cm 0cm]{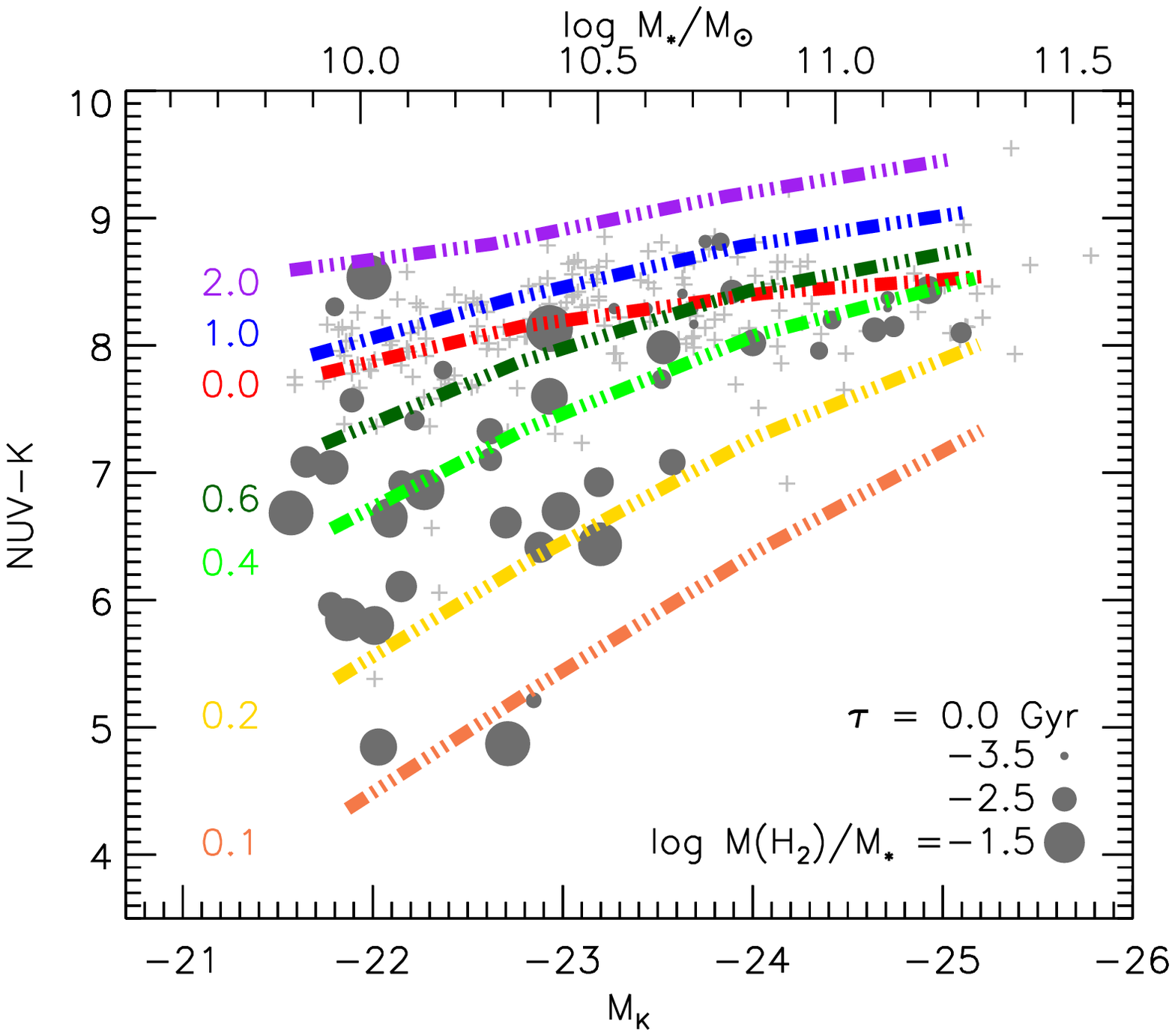}
\includegraphics[scale=0.6,clip,trim=3cm 0cm 0.8cm 0cm]{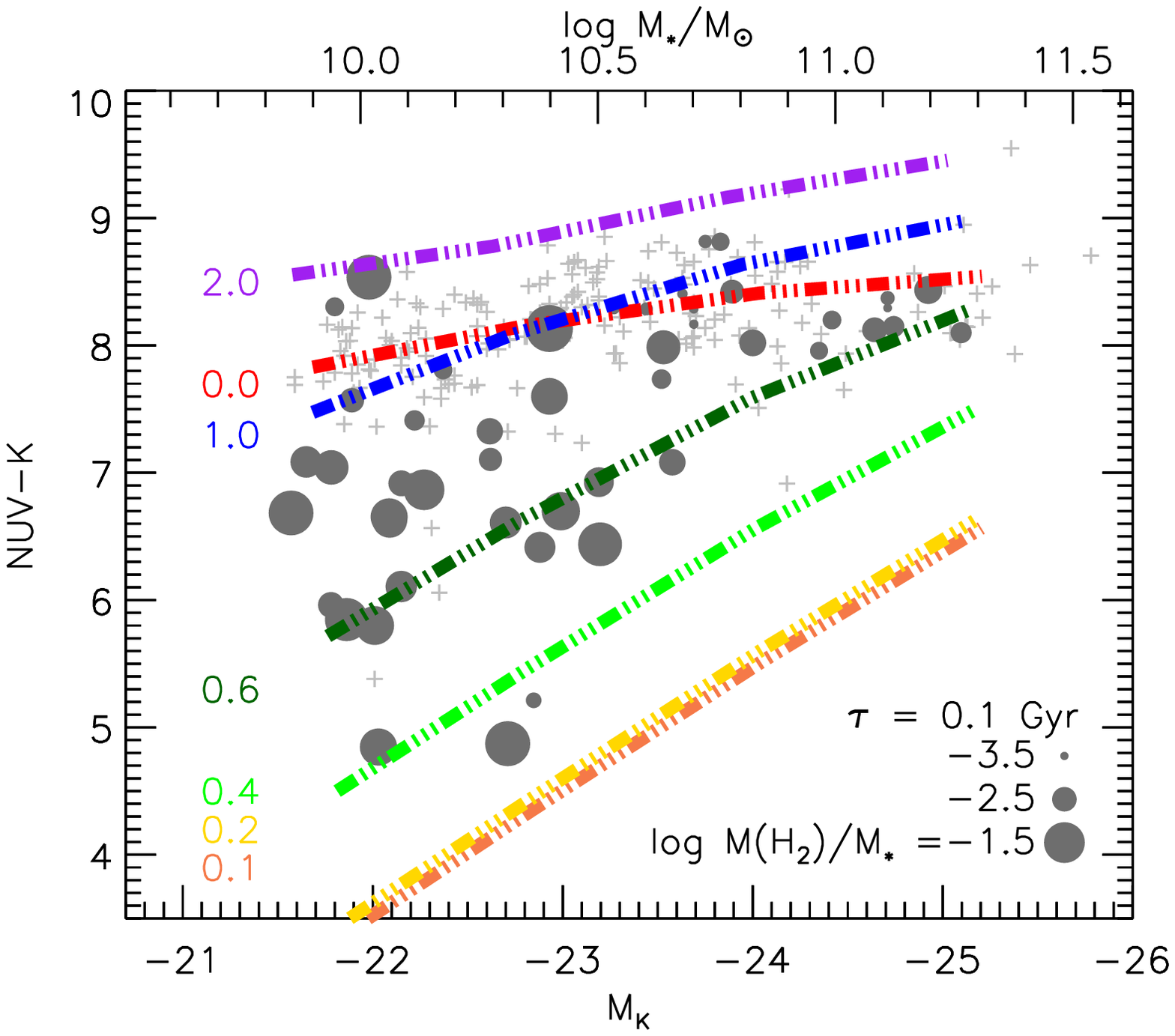}
\includegraphics[scale=0.6,trim=2cm 0cm 1cm 0cm]{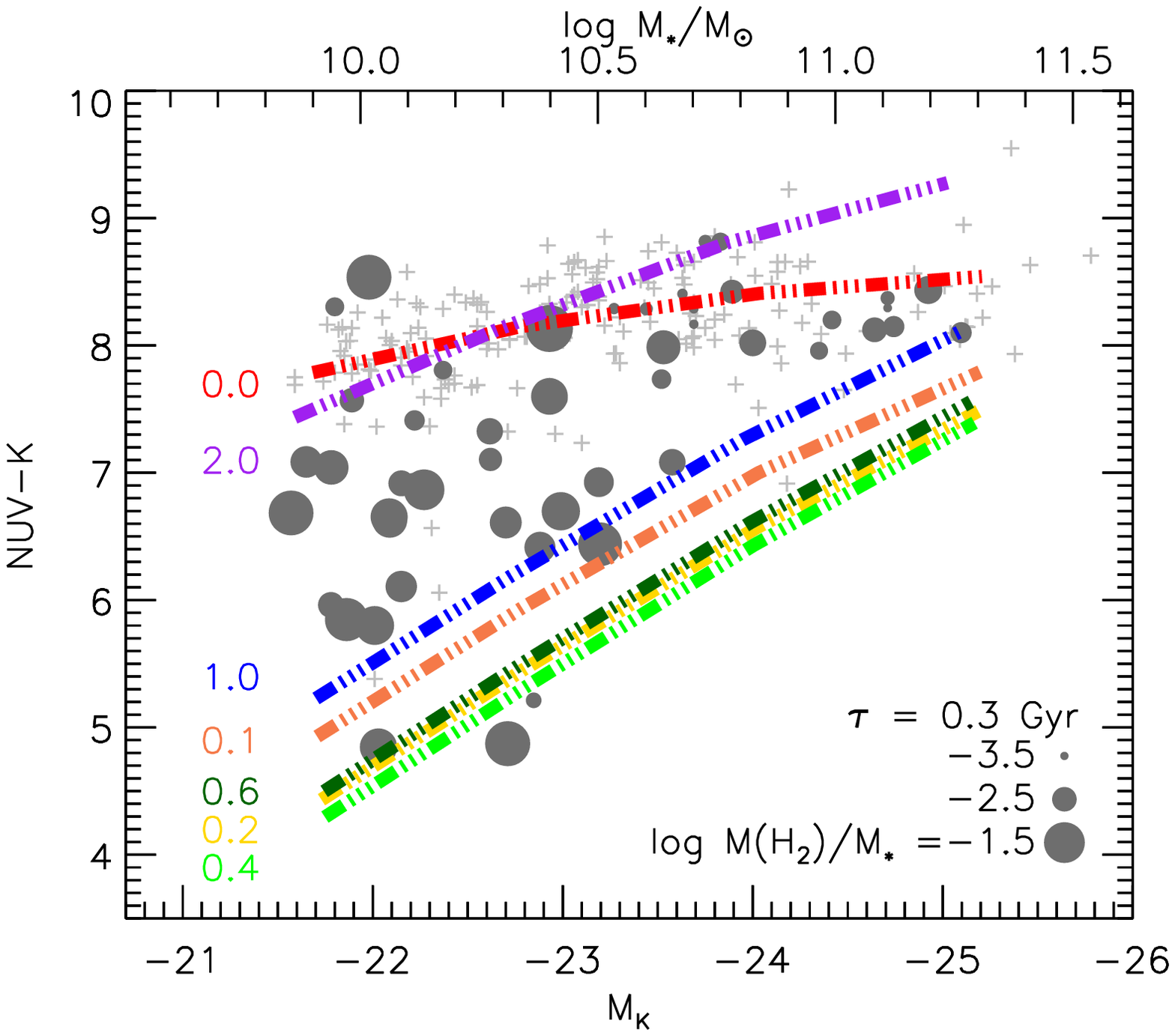}
\includegraphics[scale=0.6,clip,trim=3cm 0cm 0.8cm 0cm]{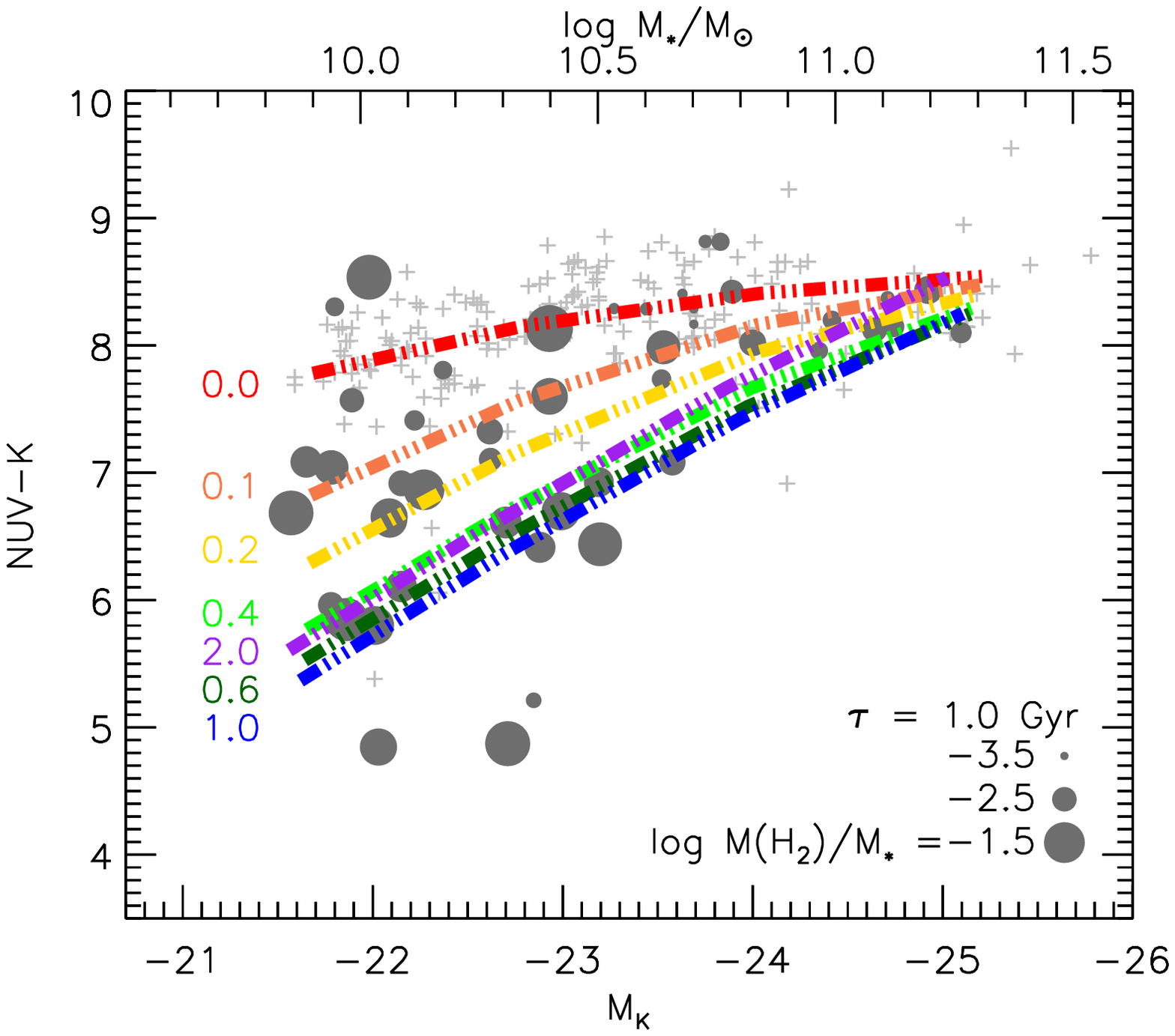}
\caption{Similar to Figure \ref{fig:galex_cmd_corr}, with \nuvk\ model isochrones.
\label{fig:toy-nuvk}}
\end{figure*}

\begin{figure*}
\includegraphics[scale=0.6,trim=2cm 0cm 1cm 0cm]{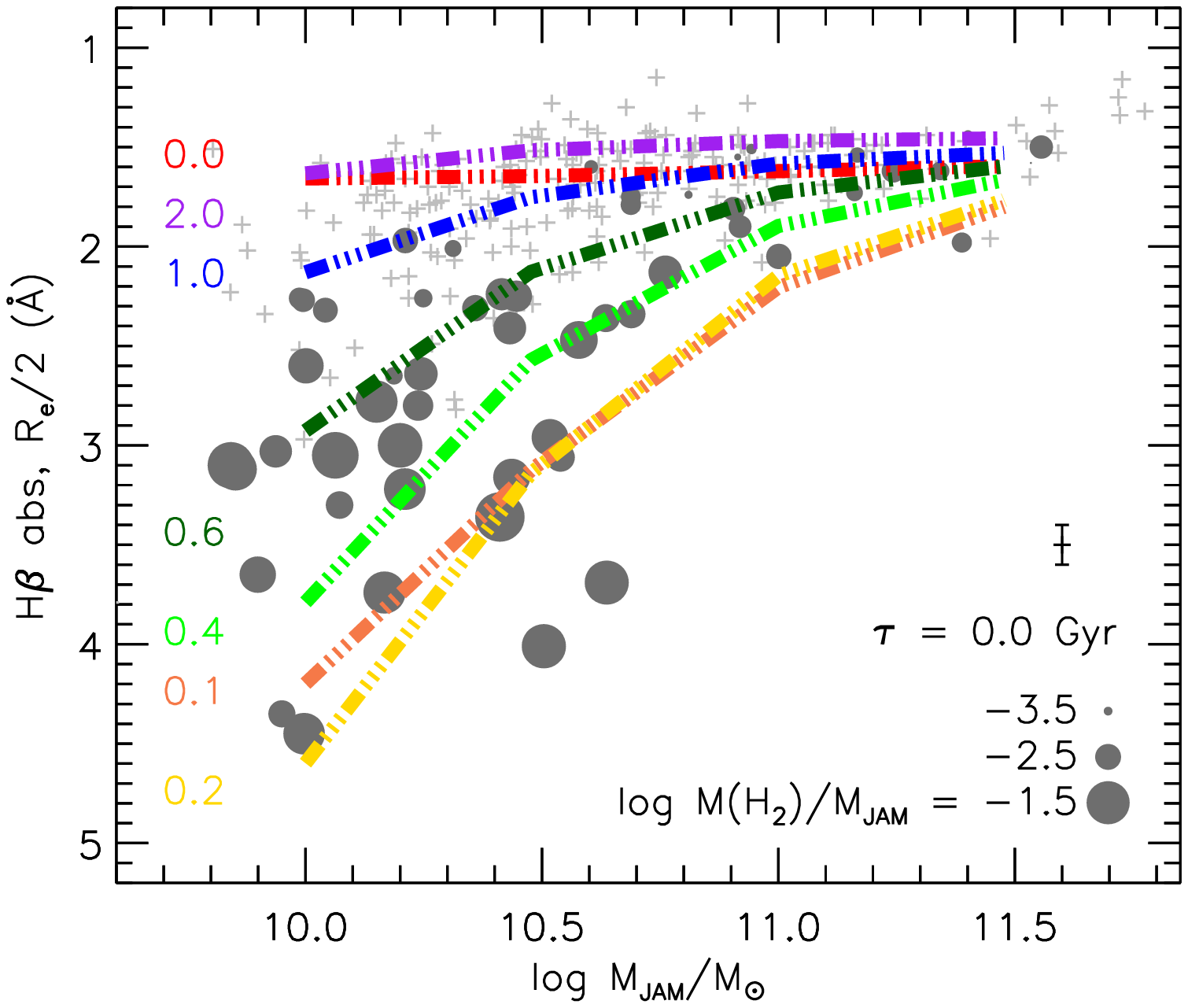}
\includegraphics[scale=0.6,clip,trim=3cm 0cm 0.8cm 0cm]{hbeta_mjam+CO+toy0.1.eps}
\includegraphics[scale=0.6,trim=2cm 0cm 1cm 0cm]{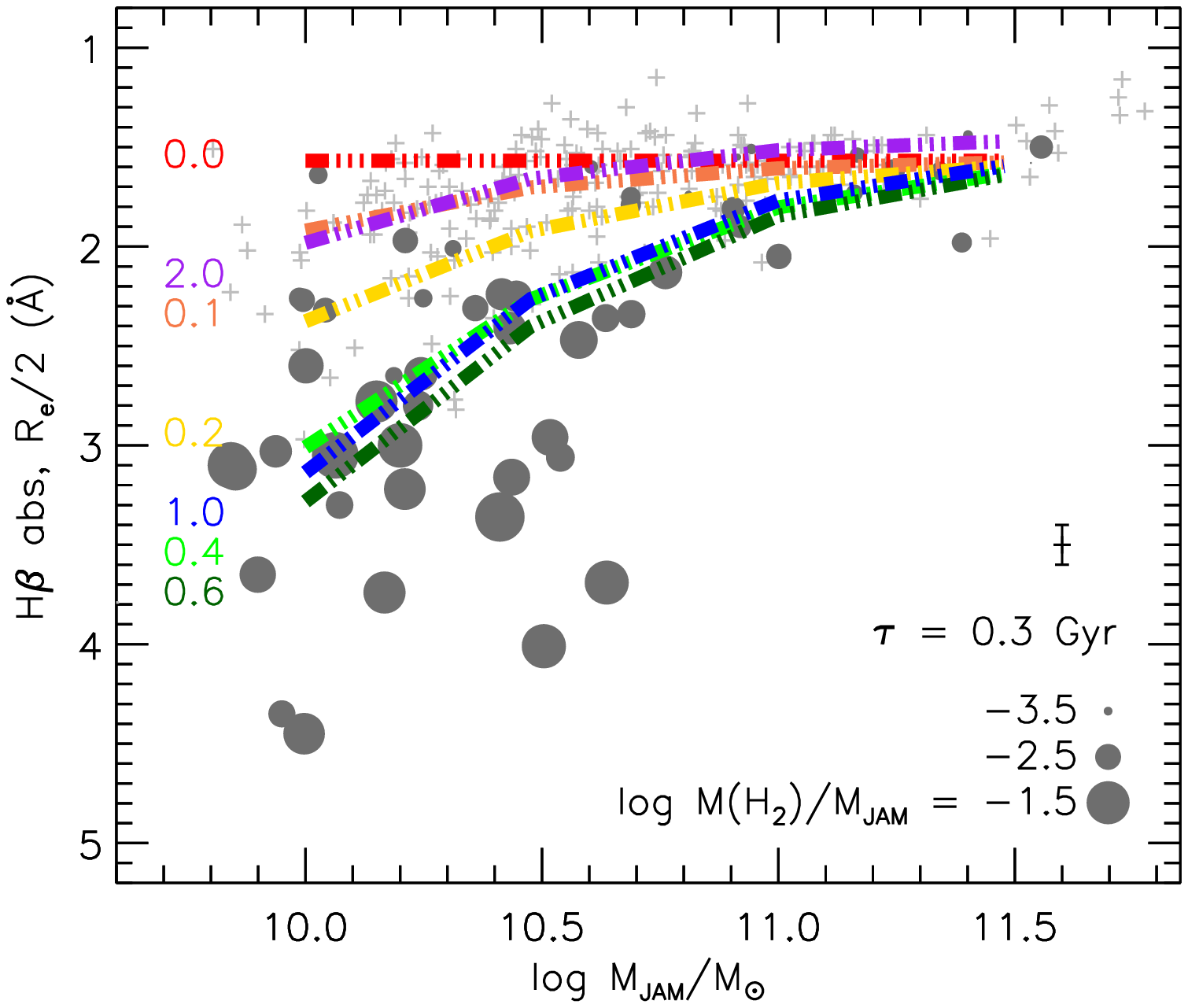}
\includegraphics[scale=0.6,clip,trim=3cm 0cm 0.8cm 0cm]{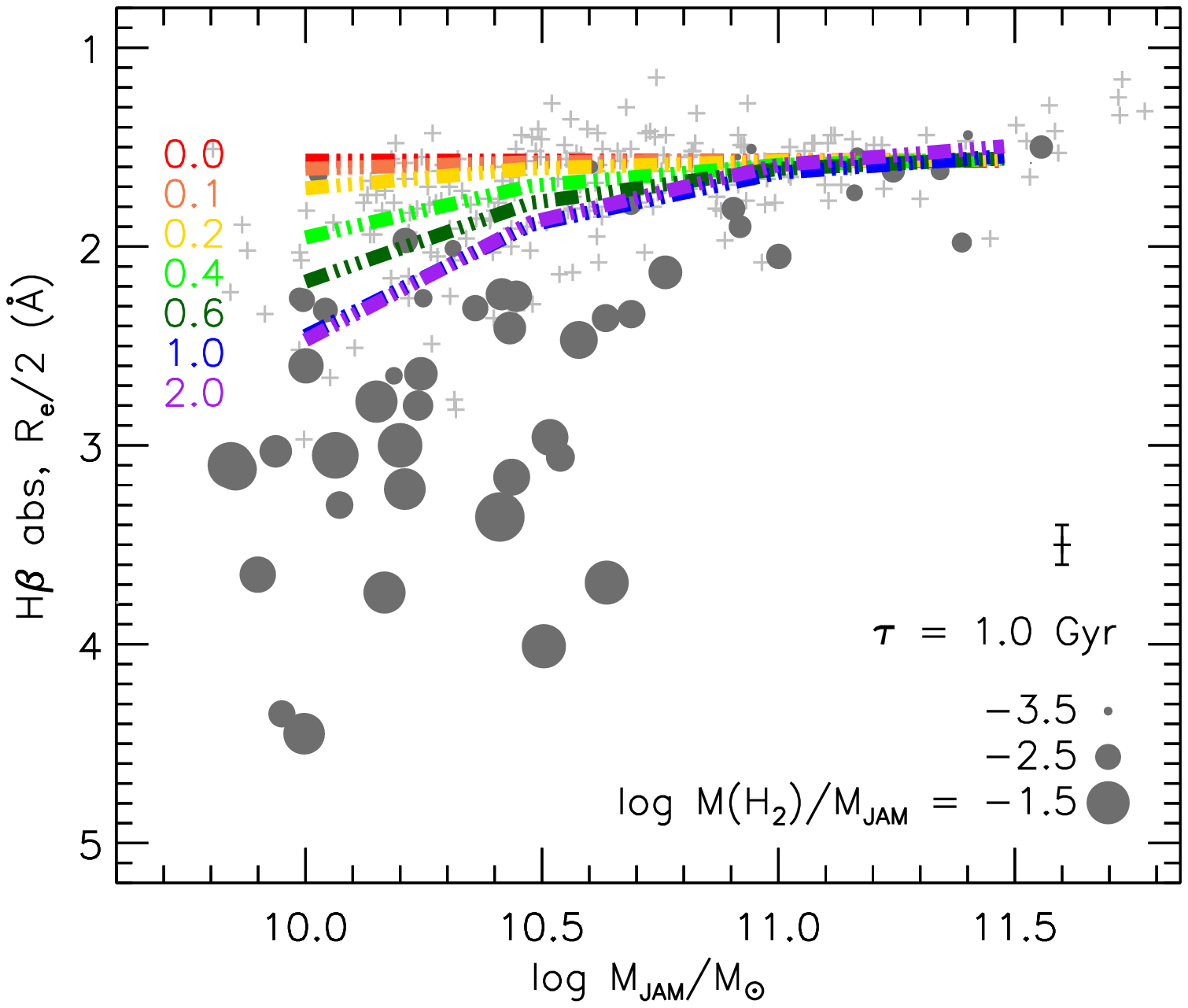}
\caption{\hbeta-mass diagram, similar to Figure \ref{fig:hbeta_cmd}, with model
isochrones. 
\label{fig:toy-hb}
}
\end{figure*}

Figures \ref{fig:toymodels} to \ref{fig:toy-hb} show the $u-r$, \nuvk, and \hbeta\
line strength models compared to the \atlas\ galaxies {\it after} correction for
internal extinction. 
As expected, the models all show an excursion off the red sequence to bluer colours
or stronger \hbeta\ absorption, followed by a gradual return to the red sequence as star formation
activity decreases.  For a fixed quantity of star formation, less massive galaxies
make larger excursions.  Spreading out the star formation over a longer time-scale
produces weaker excursions but with longer durations.  In addition,
the \nuvk\ colours are the most sensitive to the current star formation
rate. The \hbeta\ absorption line strength has a delayed response to star formation 
as it is dominated by the A stars; \hbeta\ always reaches its maximum a few
hundred Myr after $u-r$ and \nuvk\ have passed their minima.  

The slow star formation models ($\tau = 1$~Gyr, or peak SFR $<$ 0.1
\solmass\peryr) do not become blue enough, particularly in $u-r$ and \hbeta\
absorption, to explain the observed \htoo-rich blue tail galaxies.
Models with faster bursts can easily explain the blue tail galaxies; on the other
hand, such models exhibit much larger ranges of \nuvk\ colours than do the high-mass
\atlas\ galaxies.
Thus, in the context of these excursion models it is nontrivial to
explain simultaneously the tightness of the red sequence at high masses and the
numbers of blue galaxies at low masses.
In order to reproduce these effects with the simple toy models, 
we would be driven
to three possible interpretations:
{\it(i)} that the high-mass galaxies favor longer star formation time-scales 
and lower star formation efficiencies
than  the low-mass galaxies; {\it(ii)} that the high-mass galaxies had their late
burst systematically longer in the past;
and/or {\it(iii)} that the excursion model does not apply to both high-mass and
low-mass galaxies, as perhaps the low-mass galaxies
have never yet been up to the red sequence.
More detailed investigations of the galaxies' star formation histories should help
to distinguish between these possibilities.

\clearpage

\end{document}